\documentclass[Journal]{IEEEtran}
\IEEEoverridecommandlockouts
\usepackage{cite}
\usepackage{amsmath,amssymb,amsfonts}
\usepackage{algorithmic}
\usepackage{graphicx}
\usepackage{textcomp}
\usepackage{xcolor}
\usepackage{booktabs}
\usepackage{listings}
\usepackage{comment}
\usepackage{float}
\usepackage{array}
\usepackage{url}
\usepackage[hidelinks]{hyperref}

\def\BibTeX{{\rm B\kern-.05em{\sc i\kern-.025em b}\kern-.08em
    T\kern-.1667em\lower.7ex\hbox{E}\kern-.125emX}}
\begin{document}

\title{Automated Synthesis of Deterministic Cross-Domain Interfaces\\
}




\author{Konstantinos Christodoulopoulos and Antonis Selentis-Boulntadakis%
\thanks{The authors are with the Department of Informatics and
Telecommunications, National and Kapodistrian University of Athens, Athens,
Greece (e-mail: kchristodou@di.uoa.gr; aselentis@di.uoa.gr). %

A shorter version of this work has been submitted for consideration to the
\textit{IEEE Transactions on Network and Service Management}.}}
\maketitle

\begin{abstract}
Deterministic networking spans heterogeneous domains. At each boundary, two domains must agree on an assume--guarantee contract: what traffic the client may inject, and the QoS the carrier will hold for it. Composing such contracts into an end-to-end guarantee is standardized, but deriving each domain's contract is not. Today they are hand-crafted, static, and over-provisioned. The difficulty rises when traffic changes and the contract must become dynamic. We present a framework that automatically synthesizes the per-domain contract for both static and dynamic classes, along with the registration the dynamic one rests on. Two reasoning modules implemented with large language models (LLMs) drive it: an agent takes the client's traffic declaration and searches the carrier's configuration mechanisms, and a handler builds the network model from a typed disclosure of the substrate. The handler calls a network-calculus kernel for the model's hard terms, and an independent oracle---a faithful simulator, testbed, or live network---which verifies each candidate and discovers what lacks an a-priori algebraic form: when a reconfiguration is safe, and the instant to apply it. We synthesized a dynamic contract for uplink 5G fronthaul over a TDM-PON, grounded against a packet-level simulator. Across six draws from two LLM families, every synthesis produced a feasible, verified contract tight to ${\sim}1.2\,\mu$s, holding a $100$-$\mu$s deadline that reactive scheduling cannot meet, at up to $3.5$ times the bandwidth efficiency of static over-provisioning. Tasked instead with computing the worst-case delay directly, the LLMs were unsound in five of six attempts---evidence for the division of labor: LLMs construct the model, formal tools hold numeric authority. The same framework, unchanged, synthesized a static 5G--TSN bridge contract on a second substrate.
\end{abstract}

\begin{IEEEkeywords}
assume--guarantee contracts, deterministic networking, interface synthesis, large language models, network automation, network calculus, passive optical networks, quality of service, time-sensitive networking
\end{IEEEkeywords}

\section{Introduction}
\label{sec:intro}

The Internet connects independently administered networks through the IP protocol. A network domain may run its own quality-of-service (QoS) mechanisms internally, yet end-to-end the Internet is \emph{best-effort}: packets may be delayed or dropped, with no throughput guarantee. A growing range of applications---immersive media, industrial automation, smart infrastructure---carries strict QoS requirements~\cite{cps-req} that hold within a domain but break down once traffic crosses domains. The dependence runs both ways: an Internet that delivered deterministic performance end-to-end would enable applications hard to imagine today.

Deterministic Networking (DetNet)~\cite{detnet-arch} is the IETF effort to carry QoS guarantees across heterogeneous domains. A time-sensitive flow may begin at a user terminal, cross a 5G radio link and a time-sensitive networking (TSN) Ethernet segment, and traverse a metro transport before reaching a data-centre server---a chain of domains, none owning the next, extending to the compute resources at both ends. At each domain boundary, the two domains agree on what traffic may be injected and what is held for it in return: a rate, latency, jitter, or loss target, or a combination. That agreement is an \emph{assume--guarantee contract}, and it lets independent domains compose into an end-to-end guarantee.

DetNet standardizes the \emph{composition} of these contracts: it assumes each domain offers a per-domain guarantee and composes them into an end-to-end one~\cite{detnet-bl}. The single-domain mechanisms sit at layer~2: IEEE TSN, the 3GPP 5G system, and time-division-multiplexed passive optical networks (TDM-PON). Each domain rests on a shared time reference, maintaining internal synchronization and exposing (generalized) Precision Time Protocol (gPTP) at its boundaries~\cite{gptp}. Synchronization across the boundary is a precondition for any bounded QoS. What DetNet takes as given is the per-domain contract itself. Today that contract is made by hand: an engineer fixes the \emph{assume}---a static envelope the \emph{client} domain will not exceed---and provisions the \emph{carrier} domain conservatively to hold the requirement under every condition~\cite{codba}. As converged architectures multiply these boundaries---5G x-haul over optical access, TSN over 5G---the hand derivation becomes the bottleneck, and its conservative provisioning becomes permanent over-provisioning. The operator community frames the same problem as \emph{slicing}. A network slice is an end-to-end logical network carrying its own service-level agreement (SLA), and realizing one means decomposing that agreement into commitments each segment underwrites~\cite{slice-ref}. The segments are exactly these domains: a 5G slice already spans a RAN and a core, and one crossing an optical access network or reaching an industrial site spans more. Slicing standards specify how a slice is described, admitted, and orchestrated. What a segment must be configured to, so that its share of the agreement holds, is left where DetNet leaves it.

Producing the contract takes work on both sides. The \emph{client}---the domain before the boundary---supplies the \emph{assume}: a reference traffic shape for whatever produces and forwards the traffic, carrying its own uncertainties. The \emph{carrier} supplies the \emph{guarantee}. What makes a domain the carrier is that it holds the requirement from a fixed \emph{ingress} to a fixed \emph{egress}. Either may sit inside the domain, if it is first or last in the chain.
Within the carrier, a flow crosses a chain of network elements from ingress to egress, and each contributes to the end-to-end metric: it adds delay, it
constrains throughput, and it may drop packets. In typical store-and-forward
domains, what the carrier can actually set is the layer-2 forwarding mechanism at each element---a FIFO, a priority scheme, or a scheduler---which decides how packets leave the queues, through a set of \emph{control points}. A
\emph{configuration} is a choice at these control points, and it is what the
carrier sets to meet the contract.

Network calculus (NC) is the established tool for checking whether a worst-case QoS bound holds, given a forwarding mechanism, a configuration, and an arrival shape~\cite{nc}. Yet a computed bound only \emph{checks} a configuration. It does not produce one. A calculus engine, a faithful packet-level simulator, or the live network can each verify that a \emph{candidate} meets its contract. None \emph{synthesizes} the configuration that meets the target. The difficulty compounds when the admitted traffic is not one profile but several, or changes over time. A static envelope can still hold, but only by over-provisioning for the worst of them at all times; avoiding that permanent cost is what makes the contract \emph{dynamic}.

Cross-domain interfaces---the control channels that implement contracts---exist today in two classes. \emph{Static} interfaces---provisioning-time agreements, fixed when a flow is subscribed and revisited rarely---are established practice and comparatively easy to build. The hard class is \emph{dynamic} (which the standards sometimes call \emph{real-time}): the interface carries per-event contracts at the traffic's own timescale, letting the carrier follow traffic that changes faster than provisioning cycles. Even a dynamic interface has an initial \emph{static step}, a registration that admits the flow and fixes the parameters under which it is served---the standing agreement under which contracts are then minted \emph{live}. What makes a client eligible for a dynamic contract is that it knows its own traffic ahead of time. Few applications are written that way today, because no interface rewards it — a datacenter job alternating compute and communication phases, or an industrial cell reconfigured between production runs, could each declare their changes if a carrier could act on them. 

Where the dynamic class is most needed, an interface has to be standardized one domain pair at a time, and few exist because each is hard to make. The cooperative transport interface (CTI)~\cite{oran-cti} is the leading
example, and the only dynamic cross-domain interface in deployment we are aware of: it lets a TDM-PON, among others, carry 5G fronthaul efficiently, within a bounded latency. It took years to standardize, and it does not travel---a 6G numerology, or a PON whose behavior departs from what its specification shows, needs the work done again. A dynamic contract cannot be minted by hand at run time, so an interface of this kind is required. But standardizing one for every domain pair is untenable, as guaranteed boundaries spread into industrial verticals and to the compute--network edge. We therefore set out to synthesize both classes. The dynamic one is the demanding case and our focus---the dynamic contract together with the registration it rests on---and the static class follows within the same framework: a static contract is the special case in which the client announces no change, so the regime fixed at registration stays in force. 
We synthesize one interface of each class in this paper: a dynamic interface---the CTI---for the 5G fronthaul-over-PON case above, and a static interface for the 5G--TSN bridge~\cite{5gtsn}. Our synthesis produces point solutions too: an interface is derived against one
Card and one oracle, so different equipment yields a different interface. What
changes is the cost of deriving it.

\begin{figure}[t]
  \centering
  \includegraphics[width=\columnwidth]{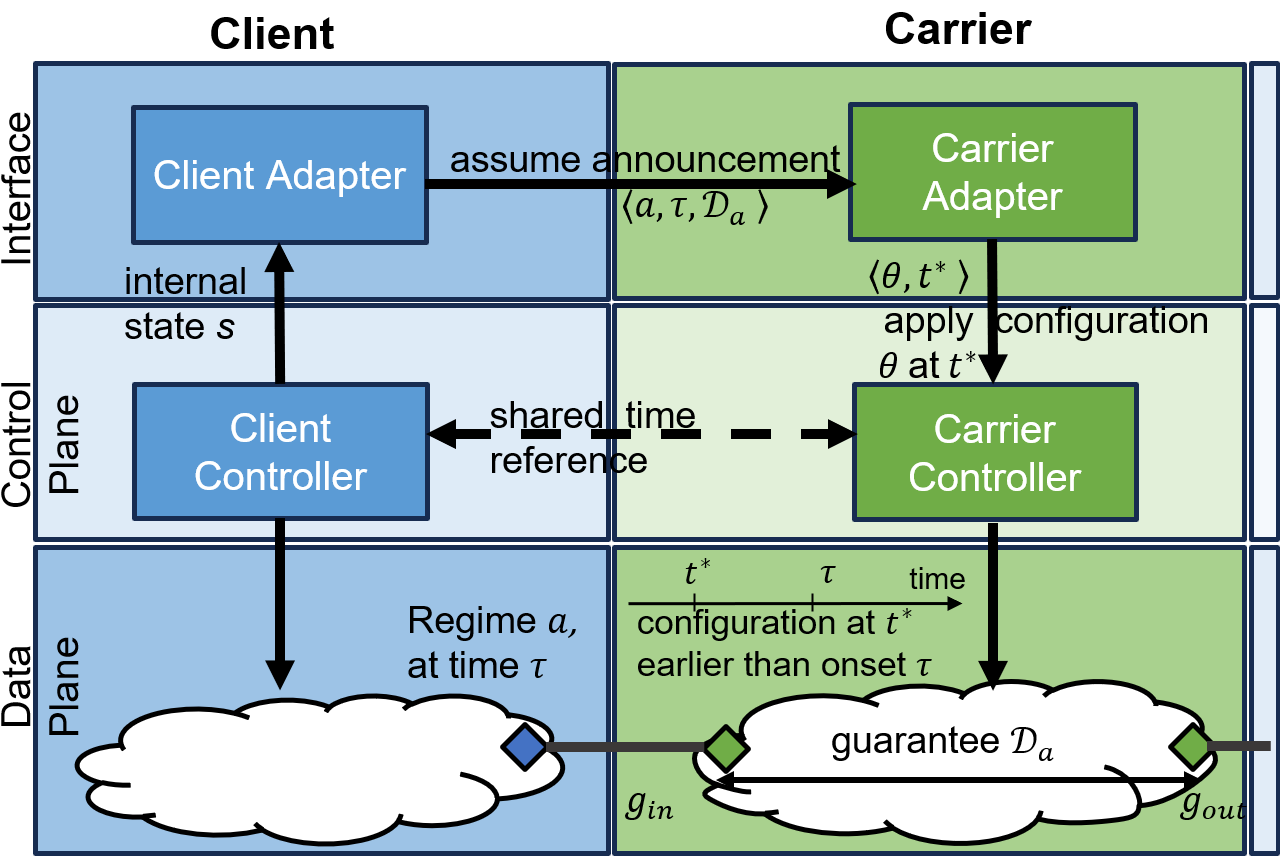}
  \caption{The \emph{online} arrangement. The synthesis that built it (Fig.~\ref{fig:synth}) has already run \emph{offline}, producing the \emph{adapters} and the \emph{contract} $\mathcal{C}$, the parametric map. In the online operation, the \emph{live} adapters enact $\mathcal{C}$ for one traffic regime: the client adapter emits the assume $\langle\alpha,\tau,\mathcal{D}_\alpha\rangle$ by translating its internal state $s$, and the carrier adapter evaluates $\mathcal{C}$ against it, passing the timed configuration $\langle\theta,t^{*}\rangle$ to the carrier's controller. In the \emph{data plane}, traffic crosses from $g_{\text{in}}$ to $g_{\text{out}}$, the span over which $\mathcal{D}_\alpha$ is held. A shared clock (gPTP) and the announcement ahead of time let the carrier apply $\theta$ before the traffic arrives ($t^{*}\!<\!\tau$, shown). Domains are generic (e.g.\ RAN over PON, TSN over 5G).}
  \label{fig:exec}
\end{figure}

We propose a framework that produces these contracts automatically. The
synthesis runs \emph{offline}; the interface it produces runs \emph{live}. We describe first what it produces---the live interface---and then the synthesis that builds it. Figure~\ref{fig:exec} shows the synthesized interface in \emph{live}
operation. Each domain runs an \emph{adapter}, and together they
hold the contract $\mathcal{C}$: the parametric map that binds the client's
assume $\langle\alpha,\tau,\mathcal{D}_\alpha\rangle$---a traffic regime $\alpha$ from the
declared set, the onset $\tau$ at which it takes effect, and the QoS
requirement set $\mathcal{D}_\alpha$---to a \emph{timed configuration} $\langle\theta,t^{*}\rangle$. 
The carrier's \emph{controller} applies $\theta$ at the disposition instant
$t^{*}$ and the \emph{data plane} then holds $\mathcal{D}_\alpha$ over the span the traffic crosses, from a fixed
ingress $g_{\text{in}}$ to a fixed egress $g_{\text{out}}$. Note that the requirement set $\mathcal{D}_\alpha$ can include one entry per guaranteed attribute in general, a single deadline in both use cases we report.
The two domains share a time reference (gPTP), so $\tau$
and $t^{*}$ lie on one clock, and its bounded inaccuracy is accounted for in
the synthesis. The shared clock and the client's advance announcement are
what make the control \emph{proactive}: the change is scheduled against the
announced onset rather than triggered by observing the traffic.

We turn to the offline synthesis that produced this arrangement. The interface is one-way, matching DetNet's one-way service: the client declares, the carrier guarantees. The client's role is therefore the lighter one. Rather than expose its proprietary internals, it translates them into a traffic \emph{declaration} in a shared language---not a single envelope but the whole regime set $A$, each regime with the requirement set $\mathcal{D}_\alpha$ it must be served under, together with the lead $\lambda_c$ by which it commits to announce each change. That shared language is what makes the framework portable: any domain that speaks it plugs in without exposing its internals. We therefore take the client's declaration as given and focus on the carrier's synthesis, which reads the carrier's substrate from a typed \emph{Domain Card}.

Carrier synthesis is driven by two reasoning modules that call two passive tools. The carrier \emph{agent} receives the declaration, reads the Card, and searches the available \emph{mechanisms}---the discrete control interfaces the substrate exposes---calling a \emph{handler} for each. The handler builds the network model for that mechanism from the same Card and sweeps its continuous parameters. To evaluate a configuration it calls a network-calculus \emph{kernel} for the terms that depend on the layer-2 forwarding mechanism, then composes them along the span into the end-to-end metric the contract names. The composition rule is per attribute and standard: latency adds along the path, throughput takes the minimum, and each attribute has its own. Configurations that meet $\mathcal{D}_\alpha$ are kept, and the handler calls an independent \emph{oracle}---a faithful simulator, testbed, or live network---to verify the computed value on the substrate and to capture the reconfiguration transients the analytic kernel cannot see.

Both reasoning modules are implemented with large language models (LLMs). 
Asked to compute a worst-case delay directly, LLMs miss by margins large
enough to break the guarantee~\cite{tsnbench}---a failure we reproduce on our
own substrate (Section~\ref{sec:eval}). Computation is therefore reserved for
the kernel and the authority to admit for the oracle. The LLMs do what they
are suited for, translating a heterogeneous substrate into a network model
and guiding the search, and the deterministic guarantee never rests on their
arithmetic.

A dynamic contract must govern how the network evolves over time---generically, for any forwarding mechanism and any QoS metric. Here we fix the QoS metric to latency, which both our synthesis use cases require. Latency is traditionally evaluated only in steady state. But when the client's declared regime changes and the carrier re-provisions to match, the network elements pass through a transient. Their queue excursions \emph{during} that transient add up, and the sum can violate the end-to-end latency requirement $\mathcal{D}_\alpha$ that both the initial and the final steady state satisfy---a case steady-state analysis does not see. The framework therefore elevates the timing of the change into the contract: the disposition instant $t^{*}$, the edge of the \emph{safe apply set}, the instants at which the new configuration may be applied while the QoS requirement still holds. Standard analysis gives no algebraic rule for this transient. We therefore developed the safety criterion while building the framework, by evaluating reconfigurations on the oracle, and formalized it: a two-window criterion with a one-sided safe set per direction of change---heavy-to-light and light-to-heavy (Appendix~\ref{app:safety}). Once established, the criterion lets the handler \emph{verify} a later reconfiguration within the declared set, instead of re-simulating it on the oracle. Proactive control depends on the client's commitment to announce each shift a lead $\lambda_c$ ahead.

Although the framework is generic across layer-2 mechanisms, slotted systems---where arrivals and service each follow their own periodic pattern of instants and sizes---are the norm for strict guarantees in TSN, 5G RAN, and TDM-PON. We therefore extend the calculus kernel for a single scheduled serving point---one slotted queue, the per-hop primitive these domains deploy. A standard rate--latency service envelope, the abstraction network calculus and DetNet typically compose~\cite{nc}, discards the fine temporal interaction between the client's slotted arrivals and the carrier's slotted service. We extended the kernel to preserve it, and return a worst-case delay that is exact where the arrival pattern and the service pattern are commensurate, stays strictly sound across the rest of the operating grid, and carries a bounded-stochastic tail (Appendix~\ref{app:comb}). The result follows from that interaction, not from a pessimistic fluid approximation.

We ground the framework in detail on one use case of each interface class, static and dynamic. Deterministic performance is usually required in both directions, and the framework applies to both. Upstream is the harder direction in access---many endpoints converging on a shared medium, served by grants---and it is the direction both use cases synthesize. The dynamic use case is uplink 5G fronthaul over TDM-PON, standardized as CTI~\cite{oran-cti}, studied in simulation~\cite{bidkar22}, and demonstrated on a testbed~\cite{bidkar23}. We re-synthesize a subset of CTI, characterize it in full, and validate it across a grid of traffic parameters against a faithful ns-3 packet-level simulator (Sections~\ref{sec:cti},~\ref{sec:eval}). The static use case is the 5G--TSN bridge~\cite{5gtsn}---a 5G system acting as a standardized TSN bridge---synthesized to its load-bearing step and validated analytically against the kernel (Section~\ref{sec:tsn}). The two use cases differ in both interface class and parameters, yet the proposed framework handles both. They are two seams of the kind a  slice is cut along---one inside a mobile network's own transport, one between a mobile network and an industrial segment---so a slice crossing either needs the contract synthesized here at that seam. Both rest on a scheduled, slotted queue---the primitive our extended kernel targets---but the framework itself is not limited to it.

This paper makes five contributions. First, we proposed a framework for the
automated synthesis of per-domain assume--guarantee contracts of both interface
classes, static and dynamic. Two reasoning modules---an agent and a handler,
both LLMs---walk the mechanism space over a typed Domain Card. The handler builds the network model for each mechanism, calls a calculus kernel for its hard terms, and calls an independent oracle to verify the result end-to-end before a contract is admitted, so numeric authority stays outside the LLMs. Second, we formalized a
reconfiguration-safety criterion that makes the timing of a change part of the
contract. We developed it on the oracle and stated it as a two-window criterion
with a directional safe set, so a later change within the declared set is
verified rather than re-simulated (Appendix~\ref{app:safety}). Third, we
extended a network-calculus kernel to slotted networks. It preserves the
temporal interaction between arrival and service slots and returns a grant-edge
and a rate-aware window delay, exact where the arrival pattern and the service pattern are commensurate
(Appendix~\ref{app:comb}). Fourth, we measured the division of labor: across six draws from two LLM families the reasoning modules reproduced
the contract's structure and its candidate ranking, whereas five of six
attempts to let an LLM compute the worst-case delay itself were unsound
(Section~\ref{sec:eval}). Fifth, we synthesized one interface of each class---
the dynamic fronthaul-over-TDM-PON contract, validated against a packet-level
simulator, and the static 5G--TSN bridge contract, validated against the
kernel---demonstrating that the construction is general
(Sections~\ref{sec:cti}--\ref{sec:tsn}).

The paper is organized as follows. Section~\ref{sec:related} positions the framework against related work. Section~\ref{sec:framework} presents the synthesis framework. Sections~\ref{sec:cti} and~\ref{sec:eval} instantiate and evaluate it on fronthaul over TDM-PON, and Section~\ref{sec:tsn} applies it to the 5G--TSN bridge. Section~\ref{sec:discussion} discusses scope and limitations, and Section~\ref{sec:conclusion} concludes.

\section{Related Work}
\label{sec:related}

The composition of per-domain deterministic contracts into an end-to-end guarantee is standardized---the synthesis of those contracts is not. We place related work by its distance from that synthesis, and locate the framework in the gap they leave.

\emph{Deterministic composition assumes the contract.} The IETF DetNet architecture and its bounded-latency model compose per-node guarantees---each node a guaranteed rate and a maximum service delay---into an end-to-end latency bound~\cite{detnet-arch,detnet-bl}. IEEE TSN and scheduled TDM-PON supply the slotted data-plane mechanisms, and 3GPP specifies the 5G system as a transparent TSN bridge that reports a bridge-delay managed object per port pair and traffic class~\cite{5gtsn}. In each case the per-domain contract is an \emph{input}. The bounded-latency model leaves the derivation of a node's rate and latency to the underlying domain, and the DetNet data model provisions flows against contracts already known~\cite{detnet-yang}. Recent data-plane simplifications such as guaranteed-latency-based forwarding still presuppose per-hop constraints~\cite{glbf}, and the reliable-wireless work addresses path redundancy rather than the derivation of a hop's base contract~\cite{raw}. The wireless and PON hops are the ones whose contract is hardest to write by hand: their per-hop service follows from the allocation a scheduler assigns, not from a datasheet, so it has to be derived against a chosen configuration. The standards leave that derivation to the operator. End-to-end network slicing states the same gap at service level: a slice's SLA is decomposed into per-segment commitments each domain must underwrite, and the frameworks specify the decomposition and the orchestration, not the derivation~\cite{slice-ref}. Today that underwriting is the same manual, conservative provisioning, which is why deterministic slices remain rare outside single-vendor deployments.

\emph{Analysis computes one envelope, and the dynamic regime was built by hand.} Network calculus is the established route to a per-node or end-to-end worst-case bound, with mature solvers behind a common interface~\cite{nc,saihu} and recent work tightening bounds for specific shapers and regulators~\cite{mohammadpour,drr-nc} and for redundancy mechanisms~\cite{pref}. Compositional performance analysis propagates standard event models---period, jitter, minimum distance---across coupled domains~\cite{cpa}. We build on this machinery rather than replace it: Nancy~\cite{nancy} is our framework's calculus engine. But these methods compute a static envelope for a \emph{fixed} configuration. They do not navigate a configuration space to emit a contract parametric in it, and they do not retain the fine interaction between slotted arrivals and slotted service together with the transient of a reconfiguration. The worst-case delay that holds in perpetuity is not the one that holds while a grant is being re-provisioned, and the latter is what a live cross-domain interface needs. Dynamic deterministic operation has been built both ways. A dynamic-deterministic network (DDN)~\cite{ddn19} re-allocated capacity at deterministic latency over optical metro hardware, co-scheduling the transport to follow the arrival. Factory-floor optical Ethernet reached a similar end while avoiding co-scheduling~\cite{jsa21}: a layer-2 tunnel that preserves packet spacing from one end to the other, carrying even the empty slots so that the timing is reproduced, and over-provisioned for the worst case. Both establish deterministic operation across a transport domain. Neither derives each domain's contract from its substrate---the step taken here.

\emph{Solvers and interface theory both presume the structure.} Where the
constraint structure is already algebraic, exact programs synthesize the
configuration itself: TSN time-aware-shaper (TAS) gate schedules are produced
by SMT and ILP over known per-hop constraints~\cite{tsn-ilp}. These are
complementary to our framework, and in its terms they are candidate tools
rather than rivals: such a program can occupy the kernel's seat, returning a
schedule that meets the requirement instead of a value to check. What it does
not supply is the constraint structure it consumes, which someone must build
from the substrate and which the handler builds here, nor a re-solve per
announced regime, since it targets a static schedule. The same holds one level
up. The assume--guarantee contract is a long-standing
abstraction~\cite{interface-theory}, expressed at many layers---the 5G--TSN
bridge-delay managed object, the DetNet rate--latency pair, and a TSN
per-stream delay bound among them---so a method that synthesizes one is
positioned to synthesize the family. Yet the automated synthesis of such a
contract \emph{across} heterogeneous domains---composing, e.g., a radio
segment's stochastic tail with a TDM-PON's grant scheduling---does not exist in
a single tool. Existing tool chains aggregate homogeneous sub-problems. The
abstraction is settled and the solvers are mature. Their cross-domain
construction is open, and it is the framework's work.

\emph{Learned methods predict and orchestrate, but do not construct the contract.} Data-driven predictors characterize 5G latency by ingesting a fixed configuration and returning a distribution or a tail estimate~\cite{det6g}. They are characterization inputs, not a parametric contract over the configuration space. LLMs have been applied to intent-based configuration and orchestration, and to early proofs of concept in TSN design flows~\cite{llm-tsn}. The decisive evidence for our architecture is negative: a recent benchmark shows LLMs asked to \emph{compute} worst-case delay against a network-calculus oracle fail by margins large enough to break the guarantee~\cite{tsnbench}---a failure we reproduce on our own substrate (Section~\ref{sec:eval}). This is precisely why the framework has the LLM \emph{construct} the network model offline and a formal kernel \emph{compute} the worst-case bound, never the reverse. The LLM shapes the model. It does not produce the number the guarantee rests on.

\emph{Position.} The framework synthesizes the per-domain assume--guarantee contract that composition assumes---automatically and parametrically, in both interface classes: the dynamic contract and the static, registration-time one. The synthesis is automatic: an agent obtains the client's description and walks the carrier's mechanism space, and for each mechanism a handler constructs the network model from a typed Domain Card and calls two passive tools---an extended network-calculus kernel that returns a worst-case bound on the QoS metric at each controlled point, which the handler composes along the span and compares against the requirement, and an independent oracle that verifies it against the substrate and fixes the disposition instant $t^{*}$. The instant of each reconfiguration is thus part of the contract, not left to the implementation. We ground the dynamic class on a 5G fronthaul-over-PON use case against a packet-level simulator~\cite{bidkar22}, and the static class on a 5G--TSN bridge use case---a second carrier with quite different arrival and service periods and sizes---validated analytically against the kernel, showing the construction is general rather than tuned to one substrate.


\section{Framework}
\label{sec:framework}

\subsection{Overview}
\label{subsec:overview}

\begin{figure*}[t]
  \centering
  \includegraphics[width=\textwidth]{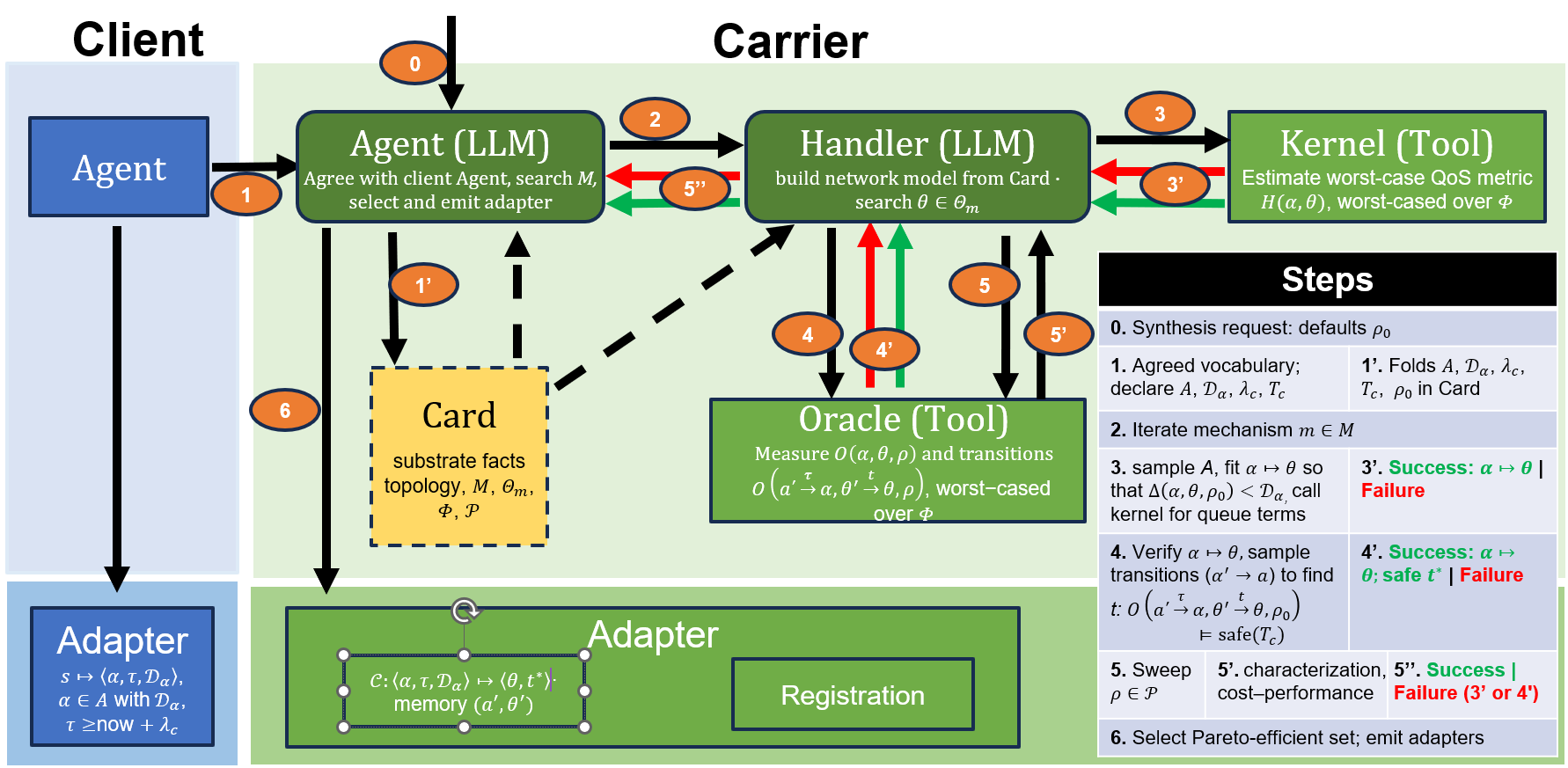}
\caption{The synthesis workflow (offline), in the six steps of
Section~\ref{subsec:overview}. It starts with a synthesis request supplying the default registration context $\rho_0$ (step~0). The client's \emph{agent} translates the space of internal states the client may reach into a \emph{declaration}---the regimes it may present and its timing
commitments---which enters the carrier's Card (step~1). On the carrier side an \emph{agent} picks a mechanism $m$ (step~2) and hands it to a \emph{handler}, which builds the network model and searches $\Theta_m$ for configurations that are steady-state feasible, calling a network-calculus \emph{kernel} $H$ for the hard terms and assembling them into QoS metric estimation $\Delta$ (step~3). The handler then drives an \emph{oracle} $O$---standing in for the live control and data planes---to verify that steady-state result on the substrate, to evaluate the reconfiguration transient the kernel cannot see, and to locate the disposition instant $t^{*}$ (step~4). A candidate failing step~3 or~4 returns the agent to step~2 for the next $m$; an admitted one is characterized on the oracle over the registration space $\mathcal{P}$ (step~5), and the agent selects among the admitted candidates (step~6). \emph{Outputs}, crossing to the online arrangement of Fig.~\ref{fig:exec}: the contract $\mathcal{C}$, the two adapters that enact it, and the registration interface with its acceptance criteria.}
  \label{fig:synth}
\end{figure*}

The framework runs in two phases, and keeping them strictly apart is essential. The synthesis phase runs \emph{offline} (Fig.~\ref{fig:synth}). A synthesis request names the domains and supplies the deployment defaults. It is raised by a human planner or by an automated process, to provision a slice or to build a link in a DetNet chain. Synthesis consumes a disclosed description of the carrier network---the carrier's Domain Card---and the client's declaration, written in a shared language the two domains agree on: the traffic regimes it may present, the requirement each must be served under, and the lead time by which it commits to announce a change (we detail later why the client's side simplifies to just this declaration). Both fold into the carrier's Card, annotated with their source, so we can speak of the Card as the single input to the carrier's synthesis. Synthesis emits the two adapters, the dynamic contract---which is a map---and a registration interface with acceptance criteria. The operation phase runs \emph{online} (Fig.~\ref{fig:exec}): the adapters implement a live interface and are activated through the registration step. They enact the contract map in real time. The client adapter translates its internal state into the next traffic regime and announces it, and the carrier adapter evaluates the map to drive the carrier's control plane and through it the data plane. Crucially, no synthesis reasoning runs in live operation. The live interface is not thereby redundant: the map is precomputed, but which regime arrives, at which onset, and under which requirement is unknown until the client announces it, so the adapters do the one thing precomputation cannot---evaluate the map for the regime the client has actually announced, and enact the configuration it returns before that regime takes effect.

Because synthesis is best understood as producing that live interface, we describe the interface first. Online, the client announces an \emph{assume} $\langle\alpha,\tau,\mathcal{D}_\alpha\rangle$: a traffic regime $\alpha$, the onset $\tau$ from which it is in force, and the QoS requirement set $\mathcal{D}_\alpha$ to be held over the carrier's span, one entry per QoS attribute in general, from a specific ingress $g_{\text{in}}$ to a specific egress $g_{\text{out}}$. The carrier answers with a \emph{timed configuration} $\langle\theta,t^{*}\rangle$: the configuration $\theta$ to apply, and the \emph{disposition instant} $t^{*}$ that fixes when. The contract $\mathcal{C}$ is the map that binds them, and Section~\ref{subsec:contract} states it formally; we introduce it here to define the symbols used thereafter. Both $\tau$ and $t^{*}$ are absolute instants on a clock the two domains share, since the carrier's controller must be told when to act.

Which side of $t^{*}$ is \emph{safe} depends on the direction of the change. A change to a \emph{heavier} regime is safe applied \emph{early}, at any $t \le t^{*}$, and a change to a \emph{lighter} one safe applied \emph{late}, at any $t \ge t^{*}$. Appendix~\ref{app:safety} defines the order and derives the asymmetry. Only the heavier direction consumes lead time. The client commits to a lead time $\lambda_c$---each new regime is declared at latest $\tau-\lambda_c$, and $\lambda_c$ absorbs the adapter's bounded evaluation time---and to a minimum spacing $T_c$ between successive onsets. Where $\lambda_c$ says how much warning the carrier gets, $T_c$ says how long it has to settle before the next change may arrive. Thus, feasibility requires $\tau-t^{*}\le\lambda_c$, the required lead within the committed one. Because the safe side is set by the direction, the carrier adapter evaluates each change against the operating point in force, $(\alpha',\theta')$, as well as the announced assume, so it holds that memory.

The rest of this section focuses on the offline synthesis that produces this interface. Synthesis uses two reasoning modules, the \emph{agent} and the \emph{handler}, and two passive tools, the \emph{kernel} and the \emph{oracle}. The kernel---a network-calculus (NC) engine---is driven by the handler to compute the hard terms of the steady-state worst-case QoS metric, for the network model the handler creates. The oracle stands in for the live control and data planes, absent during synthesis: a faithful simulator, a testbed, or, rarely, the live carrier itself. It does two jobs. It determines the dynamic behavior the steady-state model cannot express, and it verifies the synthesis before the contract is finalized, so a mistranslated network model fails against measured behavior instead of validating itself (Section~\ref{subsec:repro} measures such a catch). Soundness is therefore relative to the oracle: an oracle that diverges from the deployed carrier is a modeling error the framework cannot see, though the automated synthesis is computationally cheap to re-run against a corrected one.
 
Synthesis sees the client's declared set, not a specific announcement. The declaration gives the regime set $A$, pairing each regime $\alpha \in A$ with the QoS requirement set $\mathcal{D}_\alpha$ it must be served under---a single latency deadline for every $\alpha$ in both use cases of this paper---together with $\lambda_c$ and $T_c$. Which regime is presented, and its onset $\tau$, are known only online. Synthesis therefore computes no $t^{*}$ of its own. It reasons over \emph{transitions}, ordered pairs $\alpha' \to \alpha$ drawn from $A$, and determines for each direction (heavier or lighter) how far ahead of the onset the change must land. The adapter turns that into an absolute $t^{*}$ once the onset is announced.

On the carrier side, the carrier agent searches the mechanisms $M$ the carrier's Card exposes (e.g., discrete control interfaces like scheduling engines applied universally to all $\alpha$), each with its own parameter space $\Theta_m$, also declared in the Card. We call $\Theta_m$ continuous relative to the mechanism choice---it is the tuning within one control interface, though some of its parameters may themselves take discrete values. The strictly discrete axis of the search is the choice among mechanisms (Section~\ref{subsec:synthesis}). The search therefore selects a mechanism $m \in M$ and, within it, a parameter point $\theta \in \Theta_m$; the pair $(m,\theta)$ is a candidate \emph{configuration}. Because a contract is synthesized for one mechanism at a time, we write the configuration as $\theta$ wherever $m$ is fixed, and restore $(m,\theta)$ only where the mechanism varies.

A single $\theta$ need not serve the whole of $A$, and $A$ is in general not finite: in the fronthaul use case of Section~\ref{sec:cti} the regimes differ in the size of each arrival, which varies every 5G scheduling period. The handler therefore cannot enumerate $A$. It samples regimes, uses the kernel to find a configuration that serves each, and fits the mapping $\alpha \mapsto \theta$ across the set. That mapping is the core of the contract---it is what makes $\mathcal{C}$ parametric in the regime rather than a table of cases. The same limit applies to transitions, and Section~\ref{subsec:oracle} states how the handler covers them.

Besides the client's declaration ($A$, $\mathcal{D}_\alpha$, $\lambda_c$, $T_c$), two further sets of parameters are not searched but must be covered. \emph{Uncontrolled} parameters $\Phi$ are unset and adversarial: the relative clock phase, arrival jitter, bounded time-sync error, and the bounded variation of the carrier's own actuation. The synthesis envelopes and worst-cases them. The two sides supply different parts: the client's declaration provides the arrival-side uncertainty, the carrier's Card provides its own actuation variation, and the accuracy of the shared clock is a property of the boundary itself. \emph{Registration} parameters $\rho$ are fixed per deployment, among them the default ingress and egress $g_{\text{in}}$, $g_{\text{out}}$. Synthesis holds these at a default $\rho_0$ in its initial steps, and sweeps the space $\mathcal{P}$ at later steps to characterize the admitted configurations. In the operating network they are supplied at registration, before the dynamic interface goes live.

The handler translates a candidate mechanism and its parameters into the network model each tool consumes, and we fold that translation into the notation. We write $\Delta(\alpha,\theta,\rho)$ for the analytic worst-case metric over the contracted span, for regime $\alpha$ under configuration $\theta$ in registration context $\rho$, worst-cased over the uncontrolled parameters $\Phi$, and $H(\alpha,\theta)$ for a single kernel call at one controlled point. The call returns that point's hard term---the part the layer-2 forwarding mechanism sets, by deciding how packets are queued and served---together with the regime leaving the point, which is the arrival at the next one. The handler assembles $\Delta$ from the kernel's terms and the closed-form remainder, which the Card and $\rho$ fix: the registration context therefore enters $\Delta$ but never reaches $H$. Section~\ref{subsec:decomp} gives the decomposition. Where the span holds several controlled points the handler chains the calls into a single end-to-end $\Delta$, compared against $\mathcal{D}_\alpha$, so no per-element QoS metric has to be invented. The oracle returns two things: a) the same quantity measured end to end on the substrate, $O(\alpha,\theta,\rho)$, with no decomposition, and b) quantities measured across a transition,
\[
O\big(\alpha' \xrightarrow{\;\tau\;} \alpha,\;\;
      \theta' \xrightarrow{\;t\;} \theta,\;\; \rho \big),
\]
the regime changing at the assumed onset $\tau$ and the configuration at the apply instant $t$. Writing $\le$ for \emph{meets the requirement}---literal for a latency requirement, reversed for a metric such as throughput---the framework checks three things:
\begin{itemize}
\item \emph{steady state}: the model is stable and $\Delta(\alpha,\theta,\rho_0)\le\mathcal{D}_\alpha$ over the sampled regimes across $A$ and the fitted mapping;
\item \emph{safety}: for the extreme transition $\alpha'\!\to\!\alpha$ in each direction, the safe set collects the apply instants $t$ at which the transition run $O(\alpha' \xrightarrow{\;\tau\;} \alpha,\; \theta' \xrightarrow{\;t\;} \theta,\; \rho_0)$ satisfies the two-window criterion of Appendix~\ref{app:safety} at the declared spacing $T_c$, where each window is judged against the requirement of the regime it carries; $t^*$ is its edge---the last instant that is still safe applying early, the first applying late;
\item \emph{verification}: $O(\alpha,\theta,\rho)\le\mathcal{D}_\alpha$, first at $\rho_0$ against the kernel's result, then swept over $\rho\in\mathcal{P}$ to characterize the admitted configuration.
\end{itemize}

Which class (dynamic or static) the workflow synthesizes is decided by the client. Deterministic performance cannot be given to random traffic. The client can declare only one envelope — the usual case in industrial TSN, where the flow is periodic and does not change — gets a \emph{static} contract, fixed at registration. The \emph{dynamic} class asks two things more: that the states the client may reach produce regimes enumerable offline as the set $A$, and that each change be announced before it takes effect, with a committed lead $\lambda_c$ and spacing $T_c$. A 5G system where the distributed unit (DU) computes its uplink schedule before the traffic exists meets both.

Having defined the notation and outlined the modules that participate in the synthesis (summarized in Table~\ref{tab:terms}) we now outline the synthesis workflow. It starts with a synthesis request that provides default parameters $\rho_0$ (also referred to as step~0) and runs in six steps (Fig.~\ref{fig:synth}):
\begin{enumerate}
\item The client's agent declares the assume---the regimes $A$ it will present, the requirement set $\mathcal{D}_\alpha$ each carries, the lead 
time $\lambda_c$ by which each new regime is announced ahead of its onset, and the minimum spacing $T_c$ between successive onsets---by translating the client's internal state into a declaration in the shared language.
\item On the carrier, the agent picks a mechanism $m \in M$ (a control interface) and hands it to the handler.
\item The handler constructs the carrier's network model and searches $\Theta_m$ for a configuration that is steady-state feasible for each sampled regime, fitting the mapping $\alpha \mapsto \theta$ across $A$; the mechanism fails if some regime admits none. This is the steady-state check. It calls the kernel $H$ for the model's hard terms, such as the worst-case queueing delay, worst-casing over $\Phi$, and composes them at the default registration context $\rho_0$ into the end-to-end metric $\Delta$, which it compares against $\mathcal{D}_\alpha$. A mechanism whose limiting term the kernel cannot express passes this step undecided, and step~4 settles it on the oracle.
\item The handler queries the oracle $O$---which stands in for the carrier's live control and data planes---to verify the kernel's steady-state result on the substrate. It then drives the oracle through the transitions within $A$ to evaluate the dynamic behavior the kernel cannot express. Because the transient excursion has no a-priori algebraic form, the handler writes the simulation scenarios that identify $t^{*}$, the safety check above. The framework then checks that the light-to-heavy direction leaves room to act, $\tau-t^{*}\le\lambda_c$.
\item A candidate that fails either check returns to agent step 2, which advances to the next $m$. One that passes both is admitted as the timed configuration $\langle\theta,t^{*}\rangle$, and the handler completes the verification check, sweeping the registration space $\mathcal{P}$ to map its deployment envelope and its performance point.
\item Across the admitted candidates, each carrying its step-5 characterization, the agent selects---keeping the Pareto-efficient set or composing them---and the workflow emits its outputs: the contract $\mathcal{C}$, the two adapters, and the static registration interface with its acceptance criteria.
\end{enumerate}

We concentrate this section on the carrier's steps (2--6). The client's single act of declaring (step 1) is simplified, as we justify in Section~\ref{subsec:contract}. The two reasoning modules, the agent and the handler, are implemented with large language models (LLMs). There are, in fact, two agents: the client's agent declares (step 1) and takes no further part. The carrier's agent settles the shared language with it, receives the declaration into the Card, then walks the mechanisms $M$ and hands each candidate $m$ to the handler. Crucially, neither computes nor verifies the worst-case metric. LLMs have been shown to fail at computing worst-case delays directly, missing by margins large enough to break the guarantee~\cite{tsnbench}, so the handler only assembles the network model and writes the inputs to the three checks above. The authority is split accordingly: the kernel holds the numeric authority, the oracle holds the authority to admit, and the reasoning modules hold neither. Note that the dynamic contract is a strict superset of the static one, which is the same
workflow with step~4 bypassed. Section~\ref{subsec:outputs} states it, and
Section~\ref{sec:tsn} synthesizes one.

What makes the workflow generic is the level at which it reasons. It assumes only that the carrier is a DetNet domain exposing a service function: an assume--guarantee interface it must hold safely across reconfigurations~\cite{detnet-arch}. The function states what the domain guarantees, not when and how it is achieved; what the framework synthesizes is its realization on a concrete substrate. A new carrier domain is onboarded by writing its Card and pairing it with a kernel and an oracle. A calculus kernel exists for most forwarding principles, and the oracle seat accepts a simulator or a testbed, while the workflow and the reasoning components remain unchanged. The agent and the handler are not rewritten. The architecture is therefore generic in two dimensions, over the guaranteed attribute and over the carrier's service model, and Section~\ref{subsec:decomp} specializes both where our use cases force it. This is also where the framework meets slicing. A slice is realized as a chain of such domains, and its agreement holds only if each segment's commitment is derived and then held across reconfiguration. The contract synthesized here is that commitment at one seam, so a slice spanning a radio, a transport, and a core segment calls the workflow once per seam rather than deriving the chain as a whole.

\begin{table}[t]
\caption{Terms and notation. Upper block: the synthesis (offline) and
operation (online) components. Lower block: notation.}
\label{tab:terms}
\centering
\begin{tabular}{p{0.20\columnwidth}p{0.70\columnwidth}}
\hline
\textbf{Term} & \textbf{Scope \& Definition} \\
\hline
\multicolumn{2}{l}{\emph{Components (offline synthesis, online operation)}} \\
Domain Card & typed substrate parameters, knobs, and topology, each annotated
with its source \\
Agent & offline; two instances: the client's declares (step 1); the carrier's
walks the mechanisms $M$, advances on failure, and selects across admitted
candidates (step 6) \\
Handler & offline; builds the network model, searches $\Theta_m$ and fits
$\alpha \mapsto \theta$; drives the kernel and the oracle; returns a
configuration or failure \\
Kernel & offline; network-calculus engine; computes the configuration-dependent term at each controlled point---the part the forwarding mechanism sets, by deciding how packets are queued and served \\
Oracle & offline; stands in for the live control and data planes
(simulator/testbed/live network); verifies steady state, evaluates transients, identifies $t^{*}$ (step 4), and sweeps $\mathcal{P}$ (step 5) \\
Adapter & online; synthesis output; software enacting the contract
$\mathcal{C}$ during live operation \\
\hline
\multicolumn{2}{l}{\emph{Variables and spaces}} \\
$M$, $\Theta_m$ & mechanisms (control interfaces) and their parameter spaces; a
candidate configuration is the pair $(m,\theta)$ with $\theta \in \Theta_m$,
written $\theta$ once $m$ is fixed \\
$A$, $\mathcal{D}_\alpha$ & declared regime set: arrival pattern with parameter bounds,
in general not finite; each regime $\alpha \in A$ carries the requirement set
$\mathcal{D}_\alpha$ it must be served under \\
$\langle\alpha, \tau, \mathcal{D}_\alpha\rangle$ & the assume, announced online: regime
$\alpha \in A$, onset $\tau$, requirement set $\mathcal{D}_\alpha$; it arrives as a transition from
the previous regime $\alpha'$, in force since $\tau'$ \\
$\lambda_c$, $T_c$ & the client's timing commitments: the announcement lead time,
net of the adapter's evaluation time (each regime declared at latest
$\tau-\lambda_c$), and the minimum spacing between successive onsets \\
$\Phi$ & uncontrolled parameters: relative clock phase, arrival jitter, bounded
clock error, bounded actuation variation \\
$\langle\theta, t^{*}\rangle$ & the timed configuration: configuration $\theta$
and the disposition instant $t^{*}$, the edge of the safe apply set
($t\le t^{*}$ early or $t\ge t^{*}$ late, by direction); feasibility in the
lead-consuming direction: $\tau-t^{*}\le\lambda_{c}$ \\
$\rho \in \mathcal{P}$ & registration context (default $\rho_0$), fixing
$g_{\text{in}}$, $g_{\text{out}}$ and the reach between them among others\\
$\Delta$, $H$, $O$ & worst-case metric over the contracted span: assembled by the handler from the closed-form terms and the kernel's; the kernel's hard term at one controlled point; measured on the oracle \\
$\mathcal{C}$ & the contract, the map
$\langle\alpha, \tau, \mathcal{D}_\alpha\rangle \mapsto \langle\theta, t^{*}\rangle$ \\
$g_{\text{in}}$, $g_{\text{out}}$ & carrier ingress and egress; the requirement set is held over the span between them \\
\hline
\end{tabular}
\end{table}

\subsection{The inter-domain contract}
\label{subsec:contract}

At a boundary between two domains, the framework synthesizes a dynamic contract, stated as an assume--guarantee pair~\cite{contracts}. The contract binds each side to one obligation, and each side's adapter realizes its obligation as a map. The client's obligation is the \emph{assume}. Offline, its agent works over the space $S$ of internal states the client may reach and declares the regime set $A$ those states can produce. A state $s \in S$ pairs the client's own configuration with the traffic it receives, the same pairing a kernel call reads at one element (Section~\ref{subsec:decomp}). Online, its adapter announces each change before it takes effect. When the client's state will become $s$ at the onset $\tau$, the adapter emits the assume that state produces, $s \mapsto \langle\alpha,\tau,\mathcal{D}_\alpha\rangle$, within the constraints the declaration fixed: the regime $\alpha$ lies in $A$, the onset is at least a lead $\lambda_c$ ahead, successive onsets are at least $T_c$ apart, and the requirement set is the one that regime carries $\mathcal{D}_\alpha$. In our 5G fronthaul use case $\mathcal{D}_\alpha$ holds a single latency deadline for every $\alpha$, fixed at registration by the client's equipment class, so the announcement carries the same $\mathcal{D}_\alpha$ every time. 

The carrier's obligation is the \emph{guarantee}: for any announcement meeting those constraints, hold $\mathcal{D}_\alpha$ over the span. Its adapter realizes this as the contract map
\begin{equation}
\mathcal{C} : \langle\alpha, \tau, \mathcal{D}_\alpha\rangle \mapsto
\langle\theta, t^{*}\rangle,
\label{eq:contract}
\end{equation}
parametric in the announced regime $\alpha$---the configuration $\theta$ that meets the requirement, and the disposition instant $t^{*}$ at which to apply it, the edge of the safe apply set. The map also reads the operating point in force, $(\alpha',\theta')$, because the direction of the traffic change selects which side of $t^{*}$ is safe, and the carrier adapter holds that memory. Traditionally the interface is static in both cases: fixed by hand at registration, provisioning for the declared traffic when that traffic is static, and for the worst regime it may reach when it is not. The framework instead constructs $\mathcal{C}$ offline, from the disclosed Domain Card and for generic registration parameters.

As in IETF DetNet~\cite{detnet-arch}, the contract is defined strictly between a specific ingress $g_{\text{in}}$ and egress $g_{\text{out}}$ of the carrier domain, and the requirement applies to that span. The pair need not sit at the carrier's boundaries. Where the carrier is first in the chain the traffic originates inside it, so $g_{\text{in}}$ is internal, and the regimes are declared by whatever generates the traffic there. Where the carrier is last, $g_{\text{out}}$ is internal for the same reason. They cannot both be internal, since there would be no inter-domain contract to synthesize. Where the two sit does not affect the contract: whoever delivers traffic at $g_{\text{in}}$ declares the regimes, and the carrier holds the requirement over the span between them. 

The announcement is forward-looking: the client commits to the onset at least a lead $\lambda_c$ ahead, which lets the carrier schedule the apply against $\tau$ rather than react to the traffic once it has arrived. Both use cases are read against a shared time reference, the generalized precision time protocol (gPTP) that deterministic networks maintain. That shared clock is what makes scheduling against the onset enforceable, and its bounded inaccuracy enters the synthesis among the uncontrolled parameters $\Phi$. 

The interface synthesized here belongs to the dynamic class, the harder of the two, and like any dynamic interface it rests on an initial static step. Standards place this non-runtime \emph{registration} before the interface goes live, to establish the standing agreement under which live timed configurations are then minted. In the cooperative transport interface (CTI)~\cite{oran-cti} a management plane aligns parameters when the radio unit is installed, and a TSN controller admits a flow before it runs~\cite{detnet-cp}. Registration is what fixes the deployment for a particular install: the ingress and egress pair, the reach between them, add tenants as they register, and the rest of the parameters $\rho$. The offline synthesis supplies what this step needs, the acceptance criteria for a specific deployment (step 5). Registration is thus a byproduct of synthesis rather than a separate design: a new install is admitted against a pre-computed envelope, not re-synthesized. The registration interface becomes part of the carrier's adapter. Through it, a client requesting service provides the parameters $\rho$, which the adapter stores and reads when evaluating the map.

In a chain of domains, each domain's egress is the next domain's assume, so a carrier is itself a client to the domain beyond it. Its synthesized contract is then part of the state $s$ from which it declares: what one boundary commits to deliver is what the next must assume. Synthesizing one boundary therefore supplies the input to the next, which is why the framework's unit is the per-boundary contract, and why the client's side can be taken as given here. The end-to-end guarantee follows by the composition the standards define~\cite{detnet-bl}, which consumes per-domain contracts but does not dictate how they are derived. Because each egress is the next domain's assume, the interface carries only boundary terms, so the client needs no substrate model: its adapter merely translates its state into the declared regime set $A$ (step 1). One could envision a negotiation phase at step 1. The framework treats the declaration strictly as a one-way input, and we leave negotiation for future work. The synthesis burden thus falls entirely on the carrier, which must characterize its substrate and synthesize the configuration. Which domain acts as the carrier depends on the use case: in fronthaul over PON (Section~\ref{sec:cti}) the 5G RAN is the client and the TDM-PON the carrier; in the 5G--TSN bridge (Section~\ref{sec:tsn}) the time-sensitive network is the client and the 5G system the carrier.

\subsection{Steps 1--2: the declaration and the search}
\label{subsec:synthesis}

On the client side (step 1), the client's agent translates the space of internal states it may reach---each pairing its own configuration with the traffic it receives---into the shared language. The declaration is not a single trace but the regime set $A$: the arrival's pattern and its parameter ranges, whose points are the announced regimes $\alpha$, each with the requirement $\mathcal{D}_\alpha$ it carries, and its two timing commitments: the lead time $\lambda_c$ by which each announcement precedes its onset, and the minimum time $T_c$ between successive onsets. It also includes the declared inaccuracy around them, the arrival jitter and bounded clock error that form part of the uncontrolled parameters $\Phi$. This dictates the scenarios the handler writes for the kernel and the oracle in subsequent steps, defining the client-related parameters and their swept ranges.

On the carrier side, both the agent and the handler use its Domain Card, a typed declaration of what the substrate exposes: the available mechanisms and the control points they set, and the topology---the network elements between $g_{\text{in}}$ and $g_{\text{out}}$ along with each element's parameters---from which the decomposition reads its terms (Section~\ref{subsec:decomp}). Where the substrate offers more than one path, the path is part of the registration context, so no separate symbol is needed. Both our use cases have a single path and we leave the general case for future extensions. Each Card entry carries a provenance tag, so a value from a standard, a vendor manual, or a measurement is reasoned over as a fact of known origin. Entries folded in from the synthesis request and the client's declaration carry their own provenance. A quantity the Card does not supply is reported as external rather than invented. The Card is a typed structure with a fixed set of fields, of which Section~\ref{sec:cti} gives a complete instance (Table~\ref{tab:card}); a formal grammar over those fields is future work. Because the handler is an LLM running offline over a Card of bounded size, not online during live operation, its context window and token cost when processed by the agent and handler are provisioning-time concerns rather than live constraints.

Step 2 is the carrier agent's motion: it walks the available mechanisms $M$, picks one $m$, and hands it to the handler, which takes over the parameter space $\Theta_m$. The mechanism is chosen once for the whole declared set: a candidate $m$ either
serves every regime in $A$ or the agent advances to the next. Only $\theta$
varies with the regime, through the mapping the handler fits. A failure at any subsequent check advances the agent to the next mechanism. Crucially, the handler does not retrieve a pre-stored network model for the candidate. It constructs the model directly from the Card, as step 3 develops.

\subsection{Step 3: the network model and the kernel}
\label{subsec:decomp}

In step 3, the handler constructs the network model of the carrier. Although the decomposition presented below builds delay blocks, network calculus covers all QoS attributes: a proper formal characterization of the forwarding queue bounds the worst case on throughput, loss, and jitter alongside latency~\cite{nc}. Synthesizing a contract for another attribute changes the kernel query and the composition operator, and nothing else in the workflow. Both synthesis use cases in this paper require latency, so we specialize the decomposition below to it. The steps that remain in this section---safety, characterization, and selection---read the same for any attribute. 

The handler bounds the steady-state, one-way latency---from the ingress $g_{\text{in}}$ to the egress $g_{\text{out}}$ at the default registration context $\rho_0$---across the regimes in the declared set $A$. This bound is the metric $\Delta$ of Section~\ref{subsec:overview}, and in this subsection we describe its decomposition. Between $g_{\text{in}}$ and $g_{\text{out}}$, the carrier's data plane is the sequence of network elements the Card declares (e.g., a TDM-PON has two: the optical network unit (ONU) and the optical line terminal (OLT)). Following the standard decomposition of nodal delay~\cite{kurose}, each element $b$ contributes four delays in order: processing $d^{\text{proc}}_b$, queueing $d^{\text{queue}}_b$, transmission $d^{\text{trans}}_b$ (the serialization of packets onto the output link, size divided by link rate), and propagation $d^{\text{prop}}_b$ (the link length divided by signal velocity). Note that throughout the paper, we write \emph{delay} for a per-element contribution and \emph{latency} for the end-to-end quantity the contract holds. Latency is additive along the path, so
\begin{equation}
\Delta(\alpha,\theta,\rho) = \sum_{b=1}^{N}
\left( d^{\text{proc}}_b + d^{\text{queue}}_b + d^{\text{trans}}_b +
d^{\text{prop}}_b \right),
\label{eq:decomp}
\end{equation}
where another QoS attribute would combine the per-element contributions by its own operation rather than by a sum. 

Processing, transmission, and propagation follow in closed form once the Card and the packet size are known, with $\rho$ fixing the deployment parameters they depend on, such as the reach. The kernel is strictly needed for the hard term, the queueing delay, which depends on the element's layer-2 forwarding mechanism. We write
\begin{equation}
\big(d^{\text{queue}}_b,\; \alpha_{b+1}\big) = H_b(\alpha_b,\theta),
\qquad \alpha_1 = \alpha,
\label{eq:kernelcall}
\end{equation}
for one kernel call at element $b$: it takes the regime arriving there and returns that element's bound (worst-case queueing delay) together with the regime leaving it, which is the arrival at the next element. The kernels $H_b$ can be different for different elements $b$. The handler walks the path forward from the ingress regime, and assembles $\Delta$ from the closed-form terms and the kernel's. Note that $\rho$ appears in $\Delta$ but not in $H$: the registration context shapes the closed-form terms, not the kernel's. Unlike $\rho$, the uncontrolled parameters $\Phi$ do reach $H$: the arrival jitter and the relative clock phase shape the regime the kernel sees, and the handler worst-cases over them at each call. This split in four blocks is canonical, and the kernel in use may calculate additional terms. The kernel used in Section~\ref{sec:cti}, for example, evaluates a whole-packet service curve, so $H$ returns a worst-case delay that incorporates the transmission delay alongside the queueing ($d^{\text{queue}} + d^{\text{trans}}$).

The candidate configuration $\theta$ acts directly on the forwarding mechanism, altering the queueing delay, throughput, and jitter. It is therefore the queueing term that the mechanism choice and the handler's search move, and that the contract ultimately fixes. This step is a loop driven by the handler. It authors the kernel's input, which might require translating the physical control-plane parameters of $\theta$ into the abstract mathematical parameters of the service curve, reads $H$ back, and revises the model if needed. The loop runs per regime, not once for the set. Because $A$ is in general not finite, the handler samples regimes across it and finds for each a configuration that meets the requirement,
\begin{equation}
\Delta(\alpha,\theta,\rho_0)\;\le\;\mathcal{D}_\alpha ,
\label{eq:feasible}
\end{equation}
worst-cased over $\Phi$, and fits from those samples the mapping
$\alpha \mapsto \theta$ the contract carries.

Eq.~(\ref{eq:kernelcall}) is one call at one queue, and a path may have one such queue or many. In general these queues differ, scheduled, priority, or first-come-first-served (FIFO), and may be coupled, with arrivals periodic or not. Network calculus composes them~\cite{nc}, and the forward walk above, using additions, is that composition in its simplest form. Where the kernel supports it, convolving the service curves first and taking a single bound over the composite is tighter than chaining the per-queue calls. The handler's architectural task is to build this network of service curves from the carrier's Card and let the calculus compose them for the given arrival. Where the substrate's constraints are algebraic enough for an exact
program---a constraint or integer program that schedules the whole span---it
can occupy the kernel's seat and return the composite bound directly
(Section~\ref{sec:related}). In the deterministic access carriers we study, the path typically reduces to a dominant scheduled bottleneck, such as the queue at the first network element the client attaches to, which serves as the primary control point the synthesized configuration dictates. There the sum collapses to a single $H$ and the closed-form remainder. Deterministic networks favor scheduled, slotted forwarding at such bottlenecks precisely because it yields the tightest latency control~\cite{detnet-arch}. Because standard fluid calculus discards the fine temporal interaction between slotted arrivals and slotted service, Appendix~\ref{app:comb} develops the exact form $H$ takes for this specific case, along with the stochastic correction required when the temporal grids misalign. Both use cases we synthesize fall in this category and use that form---the ONU queue against the PON's grants (Section~\ref{sec:cti}), and the UE queue against the 5G uplink (Section~\ref{sec:tsn}).

Eq.~(\ref{eq:decomp}) bounds the data-plane latency in steady state, the time for the traffic flowing through the carrier under a fixed configuration. This is the absolute limit of the formal kernel. It cannot see the dynamic transient excursion that occurs during a reconfiguration, nor can it see the control-plane actuation (setting a configuration takes time, and that time varies). Because the steady-state kernel is blind to both the actuation timing and the resulting data-plane transient, the synthesis must advance to the oracle (step 4).

\subsection{Steps 4--5: safety, timing, and characterization}
\label{subsec:oracle}

The steady-state bound the handler builds (step~3) is not taken on its word. It needs verification against a detailed data-plane evaluation that includes transients, and it says nothing about the control plane. The oracle supplies both: during synthesis it stands in for the control and data planes the adapters will drive in live operation---the arrangement of Fig.~\ref{fig:exec} minus the adapters. In step~4 the handler drives the oracle through multiple iterations to verify the model, resolve the dynamic behavior, and locate the disposition instant $t^{*}$. In step~5 it drives the oracle to characterize the solution, a candidate that has passed steps~3 and~4.

The handler's first action in step~4 is to verify the steady-state bound. It writes simulation scenarios that run the candidate configuration on the oracle, translating the same control-plane parameters it authored for the kernel. For sampled regimes $\alpha$ and the fitted $\alpha \mapsto \theta$ it checks
\begin{equation}
O(\alpha,\theta,\rho_0)\;\le\;\Delta(\alpha,\theta,\rho_0)\;\le\;
\mathcal{D}_\alpha ,
\label{eq:verify}
\end{equation}
worst-cased over $\Phi$: the measured worst-case $O$ stays under the analytic bound $\Delta$, which in turn meets the requirement $\mathcal{D}_\alpha$. The left inequality is the one that catches a mistranslated model, since a bound below the substrate's own behavior is unsound however carefully it was computed. The check is run on rejected candidates too, so that soundness is established independently of admissibility. Soundness is thus a property of the verified output rather than of the mathematical construction that proposed it (Section~\ref{sec:eval}).

Once the steady state is verified, the handler evaluates the dynamic behavior. A contract is not set once: the carrier re-provisions as demand shifts, and a reconfiguration carries a transient, an excursion in the guaranteed metric while the queue adjusts. Standard network calculus bounds steady-state envelopes but does not give this per-event transient~\cite{nc}. We extended that reasoning into the transient with the two-window criterion of Appendix~\ref{app:safety}. Beyond the steady-state admissibility of step~3, verified above, it asks two things. The \emph{bounded transient}: throughout each window, the run's worst case stays within the requirement of the regime that window carries. And \emph{return}: by the end of the window following the onset, the run's worst case is back under the destination pair's own settled value and stays there, so it is as if no transient had happened. Neither has an a-priori algebraic form. We established the criterion by observing reconfigurations on the oracle and then formalized it, so a use case now verifies it rather than re-establishes it.

Safety and timing are, in practice, one act. The handler samples transitions $\alpha' \to \alpha$ from $A$, reads the fitted mapping at both endpoints, fixes an onset $\tau$, and runs the transition on the oracle at a candidate apply instant $t$. The windows over which the run is judged are sized by the declared spacing $T_c$, since return must complete before the next change may arrive (Appendix~\ref{app:safety}). We write $\models\mathrm{safe}(T_c)$ to denote that the transition
\emph{satisfies the two-window criterion} of Appendix~\ref{app:safety}. The safe set for that transition is
\begin{equation}
\big\{\, t \;:\;
O\big(\alpha' \xrightarrow{\;\tau\;} \alpha,\;
      \theta' \xrightarrow{\;t\;} \theta,\; \rho_0\big)
\models \mathrm{safe}(T_c) \,\big\},
\label{eq:safeset}
\end{equation}
and the disposition instant $t^{*}$ is its edge: the latest safe instant where the change must land early, the earliest where it must land late. The set is one-sided, so that single instant reports it. The actuation delay of the control plane sits naturally inside this measurement, since the oracle applies each change in the carrier's own timing. In general $t^*$ depends on the transition, so what the contract carries is a mapping $(\alpha',\alpha)\mapsto t^*$, and the handler samples the transition space to fit it: the extreme transitions the set admits in each direction first, then intermediate pairs to check that none between them is later. Where the sampled values agree within a direction, as they do on both use cases here, one worst-cased $t^*$ per direction reports the mapping and no per-pair analysis is needed. If no instant is safe, the fitted configuration is too lean to absorb the transient, and the handler raises $\theta$ above the floor step~3 returned before testing again (Appendix~\ref{app:safety}). Feasibility then requires that, in the lead-consuming direction, the announcement leaves room to act by $t^{*}$: $\tau-t^{*}\le\lambda_c$. A configuration admissible in steady state can still breach the requirement if its change lands at the wrong instant, so the framework must say not only what to set but when, making the timing contractual. A configuration is feasible only when it is steady-state admissible (step~3), verified (start of step~4) and safely applicable at a feasible disposition instant (step~4). Step~4 thus admits a subset of what step~3 admits. A failure at either check returns the agent to step~2. A candidate that passes both reaches step~5.

In step~5 the handler uses the oracle to characterize the admitted candidate. The search above runs at the default registration context $\rho_0$, which fixes a nominal deployment: the ingress and egress pair, the reach between them, one attached unit, one path. The handler now sweeps the registration space $\mathcal{P}$---the install-time variables, such as the ingress--egress distance, the number of attached units, the path, and so on---to extend the contract's validity across deployment footprints without re-searching $\Theta_m$ (Section~\ref{sec:cti} runs this over the link's reach). It checks $O(\alpha,\theta,\rho) \le \mathcal{D}_\alpha$ over $\rho \in \mathcal{P}$, and the subset where the contract holds is its validity envelope $\mathcal{P}_m \subseteq \mathcal{P}$. The sweep does double duty. It produces the envelope the standards' registration consumes (Section~\ref{subsec:contract}), the acceptance test that admits a specific deployment. And it profiles the candidate---guaranteed metric, efficiency, and their behavior across the footprint---giving the cost--performance evidence the agent's final selection draws on (step~6). The split between steps~3--4 and step~5 is deliberate: the timed configuration is synthesized, and the deployment context is characterized around it, so a new installation is a fast registration against a pre-computed envelope rather than a fresh
synthesis.

\subsection{Step 6: selection and the outputs}
\label{subsec:outputs}

Step~6 closes the loop across candidates. What step~5 admits is not a single timed configuration but a synthesized contract for one mechanism: the mapping $\alpha \mapsto \theta$ fitted across $A$, the disposition instant $t^{*}$ for each direction of change, and the validity envelope $\mathcal{P}_m$ over which both hold. A live announcement draws one $\langle\theta,t^{*}\rangle$ from it, while $\mathcal{P}_m$ is for the registration phase. At step~6 the agent compares the admitted candidates on their envelopes and their cost--performance points and keeps the Pareto-efficient set. Where a single map is wanted it selects one candidate, and it can also compose several into one map. We state composition as an option and evaluate only the selection (Section~\ref{sec:eval}).

What the synthesis returns is, in general, not a single point but a family of contracts above the kernel's admissibility floor (Appendix~\ref{app:safety}). The family trades one thing against the other: provisioning further above the floor buys margin and robustness, provisioning closer to it buys bandwidth efficiency, and the contract is the region between them, feasible and safe at every point. That region is what the dynamic interface buys. A static contract over the same declared set must provision for the heaviest regime in $A$ at all times, while the dynamic one follows the regime, and holding the change at $t^{*}$ is what keeps that tracking safe at a provisioning close to the floor. The efficiency comes from tracking the traffic. The timing is what makes tracking it safe.

After step~6 the workflow emits three artifacts: the contract $\mathcal{C}$, the two adapters that enact it in live operation, and the registration interface---the parameters and acceptance criteria, for example the reach up to which the requirement is met---which is part of the carrier adapter and is consumed once per deployment. Finally, a static cross-domain interface, a static contract, is this same workflow with step~3 as the load-bearing synthesis step: the kernel's steady-state bound suffices, and the oracle is not called at step~4 for transients. Steps~5 and~6 then proceed as in the dynamic case.

\section{Synthesizing the Cooperative Transport Interface}
\label{sec:cti}

We ground the framework on a 5G fronthaul link carried over a
time-division-multiplexed passive optical network (TDM-PON), a boundary
for which a dynamic interface already exists. Uplink fronthaul traffic is known to the 5G system before it is sent: the distributed unit (DU) computes the uplink radio schedule, so it knows what uplink traffic the user equipment (UE) and thus the radio units (RU) will produce. A PON that is not told must either react to the traffic once it arrives, too slow for a fronthaul deadline, or provision for the peak at all times, which wastes capacity most of the time. The O-RAN cooperative transport interface (CTI)~\cite{oran-cti} resolves this by having the DU declare its schedule to the PON, which then provisions to anticipate it. CTI has been evaluated in simulation~\cite{bidkar22} and demonstrated~\cite{bidkar23}. It is several hundred pages of specification, the product of years of standardization.

This section synthesizes that interface from the substrate up. We do not reproduce the full standard. We target the load-bearing part, the contract itself---what the client declares, what configuration the carrier answers with, and when that configuration is applied. Knowing where the standard arrives lets us check the synthesis against a known answer, which is why we chose this use case to ground the framework in detail.

\begin{figure}[t]
  \centering
  \includegraphics[width=\columnwidth]{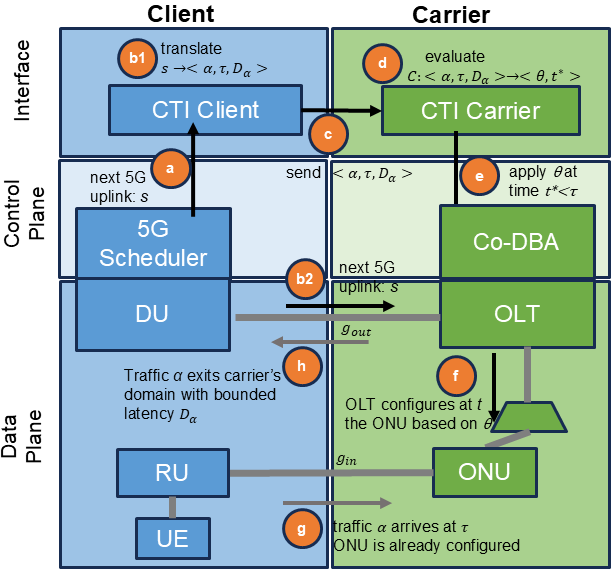}
  \caption{The fronthaul-over-PON use case in operation: the 5G system (client) carries uplink fronthaul over a TDM-PON (carrier), from the radio unit (RU) at the ONU to the distributed unit (DU) at the OLT. The lettered sequence (a)--(h) present one \emph{online} reconfiguration, beginning when the DU computes the next uplink schedule $s$. Arrow (b2) is the sole downstream arrow, carrying $s$ to the UEs as ordinary data, and is not part of the synthesized contract. Co-DBA is the carrier's mechanism, configured by $\theta$, and $g_{\text{in}}$, $g_{\text{out}}$ are the span's endpoints.}
  \label{fig:cti_op}
\end{figure}

The radio access network (RAN) is the client of the PON carrier. 5G user equipment (UE) generates traffic to a radio unit (RU), the RU attaches to a PON optical network unit (ONU), which connects via fibre to the PON optical line terminal (OLT), and the OLT in turn connects to the DU. The PON carries the fronthaul between RU and DU, and we focus on the \emph{uplink}---from the RU, through the ONU and OLT, to the DU---since strict QoS is harder to hold in this direction. The 5G system (UE, RU, and DU) is the client, the TDM-PON the carrier (Fig.~\ref{fig:cti_op}).

The ingress $g_{\text{in}}$ is the RU--ONU attachment and the egress $g_{\text{out}}$ the OLT--DU link, so both endpoints belong to the client (the 5G system). This does not change the contract: the carrier holds the latency from ingress to egress, wherever these sit (Section~\ref{subsec:contract}). The DU that computes the uplink schedule is reached through the PON. The space $S$ of client states is the schedules the DU may compute. In live operation a state $s \in S$ is a specific schedule, computed for the upcoming UE transmissions. That schedule has a dual role. First, $s$ is translated into control messages that travel downstream, together with data, from the DU to the RU and the UEs, instructing each UE when to transmit. That is why $s$ exists, and it is what makes the uplink traffic arrive at the ONU at a known time. Second, $s$ is what the client adapter translates into the assume and sends to the carrier adapter over the CTI, the dynamic interface whose adapters and mappings we synthesize. The PON and the 5G system are assumed synchronized (gPTP; both support PTPv2 in real deployments). The downstream control messages cross the PON as ordinary data the carrier does not inspect. The synthesis produces a unidirectional interface, and the one we synthesize is the upstream. The upstream is the contended direction: several ONUs share the fibre, and the OLT must arbitrate among them. The downstream QoS can use single-point scheduling methods. The control messages travel downstream under that treatment. This is why the CTI standard is defined in one direction rather than as a bidirectional interface.

The QoS requirement, a latency deadline, is tight. It depends on the 5G equipment type and is on the order of $100\,\mu$s. The obstacle CTI removes is the PON's closed control loop. To serve Internet-type applications for FTTH users a PON typically employs a dynamic bandwidth allocation (DBA) scheme that implements such a loop: the ONU reports its buffer upstream, the OLT recomputes the allocation---typically aggregating the reports of several ONUs---and issues the grants in a later bandwidth map (BWmap), after which the traffic waits in the ONU queue for its granted slot. There is usually a default grant size, and the loop absorbs traffic changes around it, but every change still travels the loop, and the latency of the arriving packets depends on it. The aggregation and recomputation run on a millisecond cycle in practice, and even with the computation neglected entirely, the round trip of carrying the observation from ONU to OLT and the grant back remains, fixed by the standard. A reactive DBA is therefore far too slow for the fronthaul deadline. CTI removes the observation leg: the DU declares its future traffic to the OLT, so the PON updates the allocation (described in the BWmap) \emph{ahead} of the traffic rather than reacting to it. In our framework's terms the mechanism is a fixed bandwidth allocation (FBA), re-provisioned from that forecast---the arrangement the standards name a cooperative DBA (Co-DBA).

The synthesis starts from three inputs: a synthesis request, the carrier's Domain Card (Table~\ref{tab:card}), and the client's declaration. At step~1 all three fold into the Card, so that every later step reads a single document. The handler's kernel is Nancy~\cite{nancy}, and its oracle is the ns-3 simulator with XG-PON extensions~\cite{xgpon-ns3} which we extended to add the FBA mechanisms and the instrumentation this evaluation needs\footnote{Our xgpon-ns3 fork: https://github.com/kchristodouUOA/xgpon-f2f}, standing in for the live control and data planes. This section details the substrate and how this interface use case maps onto the proposed framework of Section~\ref{sec:framework}. Section~\ref{sec:eval} reports the synthesis of the dynamic contract: the kernel's accuracy against the oracle, the oracle's use in locating the safe disposition instant, and the sweep of the registration parameters.

\subsection{The Card and the client's declaration}
\label{subsec:card}

The synthesis request (step~0) names the boundary and the default deployment. The RU attaches to an ONU and the output of the OLT attaches to the DU, along with fibre reach of $5$\,km by default which fixes $g_{\text{in}}$ and $g_{\text{out}}$; the direction is the uplink; and one RU--ONU pair is served as default, a single tenant on the shared medium. Together these are the default registration context $\rho_0$. The reach and the tenancy are defaults rather than fixed facts: step~5 sweeps both across $\mathcal{P}$ (Section~\ref{subsec:characterize}), and in an operating network the registration overwrites them with the installed topology.

Table~\ref{tab:card} lists the inputs. Above the rule is the carrier's Domain Card, the substrate the handler must model, and its entries group by the term of Eq.~(\ref{eq:decomp}) each one serves. The upstream line rate, the $125\,\mu$s frame period, and the frame's capacity in words and the bytes a grant may carry fix the service comb (service pattern and size). We do not carry the forward error correction (FEC) or the per-fragment overheads as separate entries, since the substrate accounts for them internally when it converts a rate request into grants. What escapes that accounting is the provisioning margin below. The fibre's propagation slope gives $d^{\text{prop}}$ directly, once $\rho_0$ fixes the reach. The Card also declares the mechanisms $M$ the substrate exposes, each with its control points and the parameter space $\Theta_m$ they span.

Here $M$ holds three control interfaces. The first is the traditional reactive DBA, realized in the oracle by the XGIANT engine~\cite{xgiant}. Its control points are the per-T-CONT (transmission-container, the upstream service entity) rate classes (fixed, assured, non-assured, and best-effort) with their service intervals. The other two are FBA interfaces, distinguished by the spacing of their grants and referred to as schemes: single-frame (FBA-SF), one grant per PON frame, so $T_{\text{ser}} = 125\,\mu$s, and quad-frame (FBA-QF), four grants per frame, so $T_{\text{ser}} = 31.25\,\mu$s. For both, the control point spanning $\Theta_m$ is the per-ONU assured information rate (AIR), from which the handler derives $G(\theta)$, the bytes granted at each service instant, which a PON reader will recognize as the upstream burst. The agent walks all three mechanisms, and Section~\ref{sec:eval} reports the outcome: FBA-SF fails the kernel's steady-state check, the reactive DBA fails at the oracle on its measured reaction floor, and FBA-QF is admitted. The path topology of Section~\ref{subsec:decomp}, the ONU and OLT, is declared in the Card, with every entry carrying a provenance tag. The Card includes values for both 25~\cite{25gspon} and 50\,Gb/s line rates: the line rate is a substrate variant, fixed at hardware initialization prior to registration, so the workflow synthesizes the interface once per rate and the registration step selects the variant the installed hardware realizes.

Below the rule is the client's declaration. It is not one traffic regime but the set $A$: the $500\,\mu$s slot pattern of fourteen arrivals, thirteen spaced $35.677\,\mu$s and one $36.198\,\mu$s, each of size $L$ bytes, and $L$ varies between schedule updates, which in the operating network are the announced regime changes. The declaration is written in the shared language, so it carries an interval and a size in bytes. On the client's side these come from the 5G numerology---the slot, the symbol duration, the physical resource blocks (PRB), the modulation format---and the client agent translates them, so that nothing 5G-specific crosses the boundary. The declaration also carries the jitter bound around those instants, part of $\Phi$, and the client's two timing commitments: the announcement lead time $\lambda_c$, four milliseconds here, and the minimum spacing $T_c$ between successive onsets, one millisecond. The requirement is a single deadline, $\mathcal{D}_\alpha = 100\,\mu$s for every $\alpha \in A$, entering through the client's equipment class and fixed at registration rather than re-declared per regime (Section~\ref{subsec:contract}). A looser class would raise $\mathcal{D}_\alpha$ and change the outcome of the steady-state check, Eq.~(\ref{eq:feasible}): a deadline above the FBA-SF bound of Section~\ref{sec:eval} would admit both FBA schemes and turn step~6 into a selection among several candidates. We keep to the single class here. At step~1 this declaration joins the PON Card, tagged with \emph{client} provenance, and the subsequent steps reason over the combined Card.

\begin{table*} [!htbp]
\centering
\caption{Inputs to synthesis for the fronthaul-over-PON use case. Above the rule: the PON Domain Card --- the carrier's substrate. Below the rule: what the synthesis request and client declares as the assume (step~1), in the shared language. At step~1 that declaration joins the Card. Every entry carries a provenance tag.}
\label{tab:card}
\small
\begin{tabular}{@{} 
  p{\dimexpr 0.25\textwidth - 1.5\tabcolsep \relax} 
  p{\dimexpr 0.39\textwidth - 1.5\tabcolsep \relax} 
  p{\dimexpr 0.36\textwidth - 1\tabcolsep \relax} @{}}
\toprule
\textbf{Fact} & \textbf{Value} & \textbf{Provenance} \\
\midrule
\multicolumn{3}{@{}l}{\textit{Line and frame}}\\
Upstream line rate & 24.8832 / 49.7664\,Gb/s & standard; substrate variant, fixed at hardware initialization; contract synthesized per rate \\
Frame period              & $125\,\mu$s                                & standard \\
Upstream frame size       & $97{,}200$ / $194{,}400$ words             & derived from line rate (standard) \\
Word size                 & $4$\,B                                     & model \emph{caveat} (hardware uses $9{,}720$ intervals and multiplier of 10/20 words per interval) \\
\addlinespace
\multicolumn{3}{@{}l}{\textit{Provisioning}}\\
Grant feasibility rule & flat $5\%$ margin above the payload rate; the substrate accounts for coding and framing internally when it converts an AIR into grants & Card entry, \emph{corrected} by measurement to the fitted edge $r \ge r_{\min}(\alpha)$, Eq.~(\ref{eq:cti-edge}) (Sec.~\ref{subsec:static}) \\
Per-grant PHY overhead (PLOu)   & 8/16 words (guard + preamble/delimiter) & standard  \\
BWmap allocation unit     & $40$\,B at $25$\,Gb/s                      & standard (C.6.1.6); $8$-B legacy pack in model \\
BWmap field width         & uint32 (in-memory)                         & model \emph{caveat} (hardware $\le 16$-bit) \\
\addlinespace
\multicolumn{3}{@{}l}{\textit{Topology and delay blocks}}\\
Path elements & ONU, OLT (single path) & Card (topology) \\
Modelled at ONU & $d^{\text{queue}}$, $d^{\text{trans}}$ (whole-packet curve: $H$ returns their sum), $d^{\text{prop}}$  & handler \\
Modelled at OLT & $d^{\text{proc}}=1.5\mu$s (FEC decode) & Card \\
Neglected & ONU $d^{\text{proc}}$; OLT $d^{\text{queue}}$, $d^{\text{trans}}$ & handler (justified, Sec.~\ref{subsec:cti-comb}) \\
\addlinespace
\multicolumn{3}{@{}l}{\textit{Fibre}}\\
Propagation slope $d^{\text{prop}}$ & $5\,\mu$s/km                     & standard (fibre) \\
Maximum reach             & $\le 10$\,km                               & substrate limit (characterized) \\
\multicolumn{3}{@{}l}{\textit{Mechanisms, configurations ($M$, $\Theta_m$)}}\\
Mechanisms $M$ & reactive DBA; FBA-SF (one grant per $125\,\mu$s frame); FBA-QF (four grants per frame, spaced $31.25\,\mu$s) & standard; vendor \\
Reactive DBA control points & XGIANT: per-T-CONT rate classes (fixed, assured, non-assured, best-effort) and service intervals & standard (T-CONT model); implementation~\cite{xgiant} \\
FBA control point & per-ONU assured information rate (AIR), spanning $\Theta_m$; handler derives the per-frame grant $G$ & vendor \\
Max TCONT per grant/ ONU/ map& $16$ / $64$ / $512$                 & standard (ns-3) \\
\midrule
\multicolumn{3}{@{}l}{\textit{Declared by the client (the assume, step~1)}}\\
Announcement lead $\lambda_c$ & $4$\,ms & client (commitment; CTI reports $4$\,ms ahead) \\
Onset spacing $T_c$ & $\ge 1$\,ms & client (commitment; CTI reports per $1$\,ms cycle) \\
Arrival pattern & $14$ per $500\,\mu$s slot: $13\times35.677\,\mu$s $+\ 1\times36.198\,\mu$s (extended CP) & client (the 5G radio slot) \\
Arrival size $L$          & $13.4$--$46.4$\,kB per arrival             & client (set by the scheduled radio resources) \\
Arrival jitter ($\Phi$)   & $\le 10\,\mu$s                             & client (uncontrolled parameter) \\
Deadline $\mathcal{D}_\alpha$           & $100\,\mu$s one-way                        & client requirement; external to the substrate~\cite{oran-fh} \\
Link reach (default $\rho_0$) & $5$\,km                                & synthesis request (deployment default); swept at step~5; overwritten at registration \\
Attached RU/ONU (default $\rho_0$) & $1$ & synthesis request (deployment default); swept at step~5; overwritten through registration \\
\bottomrule
\end{tabular}
\end{table*}

The Card and the declaration together fix the arithmetic the synthesis works in. On the client's side the arrival pattern is fixed and only the size $L(\alpha)$
varies across $A$, so a regime reduces to one variable, the offered load
$\ell(\alpha) = n_{\text{a}}L(\alpha)/T_{\text{pat}}$ in bits per second.
Writing $T_{\text{arr}} = T_{\text{pat}}/n_{\text{a}}$ for the mean arrival
period, this is $L(\alpha)/T_{\text{arr}}$, the form the expressions below use:
a rate depends on the pattern only through its mean, and the pattern's fine
structure matters where the kernel looks, not here. On the carrier's side the mechanism fixes $T_{\text{ser}}$ and the configuration fixes the grant size $G(\theta)$ through the AIR, so a configuration also reduces to one variable, the provisioned rate $p(\theta) = G(\theta)/T_{\text{ser}}$. Because both sides collapse this way, the mapping $\alpha \mapsto \theta$ becomes a relation between two rates, and we can speak of the \emph{provisioning ratio}
\begin{equation}
r = \frac{p(\theta)}{\ell(\alpha)}
  = \frac{G(\theta)}{L(\alpha)}\cdot\frac{T_{\text{arr}}}{T_{\text{ser}}} .
\label{eq:cti-ratio}
\end{equation}
This is a simplification the use case allows, not a property of the framework. In general a regime carries a pattern and a configuration several parameters, neither side collapses to a scalar, and the fitted mapping has no such closed form.

A ratio of one is not feasible. Coding and framing add to the declared payload before it occupies the line. The substrate absorbs most of that internally, since the OLT converts an AIR into grants knowing its own coding and framing, and what escapes is a residue the declaration cannot predict, because it depends on how the payload divides across grants. The handler therefore measures the least feasible grant on the oracle and fits it, in the affine form
\begin{equation}
G_{\min}(\alpha) = k\,L(\alpha) + b ,
\label{eq:cti-edge}
\end{equation}
where the slope $k$ carries the proportional share of the overhead and the fixed term $b$ the per-grant penalties. Feasibility then reads $r \ge r_{\min}(\alpha)$, with $r_{\min}$ the ratio Eq.~(\ref{eq:cti-ratio}) assigns to $G_{\min}$. Whether the fixed term matters over a given declared set is a property of the substrate and of the resolution the oracle is swept at, not of the framework: on this use case the fit came out nearly proportional, and Section~\ref{subsec:static} reports the coefficients. The Card's own entry states this margin as a flat five percent, a single constant chosen to sit above the edge everywhere rather than to track it, which is what a static format can express and what avoids fitting anything. The handler replaces it with the fitted edge, and that replacement is an instance of the $\alpha \mapsto \theta$ fitting of Section~\ref{subsec:decomp} rather than a separate step: the handler samples $A$, locates the feasible edge at each sample, and fits the relation across the set. Catching that the flat entry disagrees with the substrate takes a cross-field judgment, overhead against load, which the handler makes and the oracle confirms. Section~\ref{subsec:repro} measures it making this correction. A second, smaller catch was dimensional: an early Card draft carried a derived stability edge in mismatched units, which the handler's type check flagged, and we now keep derived quantities out of the Card.

Inverting Eq.~(\ref{eq:cti-ratio}) at a chosen $r$ gives the grant that regime needs,
\begin{equation}
G(\theta) = r\,L(\alpha)\,\frac{T_{\text{ser}}}{T_{\text{arr}}} ,
\label{eq:cti-map}
\end{equation}
which is the mapping $\alpha \mapsto \theta$ in closed form on this substrate: the declaration gives $L(\alpha)$, the chosen $r$ gives $G(\theta)$, and the AIR that yields that grant is $\theta$. Because $r$ compares two rates, one value of $r$ provisions the same capacity under either scheme. The schemes differ in how that capacity is cut into grants, $T_{\text{ser}}$ apart, which is what the beat of Section~\ref{subsec:cti-comb} then resolves. Every $r$ above the edge is a feasible mapping, and that family is what step~6 selects within, trading latency margin against the capacity the higher ratio consumes.

The framework strictly separates what the carrier \emph{can} do from what the client \emph{demands}. The deadline $\mathcal{D}_\alpha$ is a requirement external to the substrate, and it arrives strictly via the client's declaration rather than the Card. Conversely, the Card is ingested rigidly as the carrier's physical truth, typed and checked for well-formedness---and the provisioning entry above shows that rigid ingestion still leaves room for the synthesis to catch a specification that is well-formed and simply disagrees with the physics.

\subsection{Computing the delay bound via the kernel}
\label{subsec:cti-comb}

The fine temporal interaction developed in Appendix~\ref{app:comb} is made
concrete here. Each side of the queue is a \emph{comb}: a periodic train of
instants, each carrying a size. On the arrival side the pattern is the radio
slot, fourteen symbols in $T_{\text{pat}} = 500\,\mu$s; on the service side the 
grants fall every $T_{\text{ser}}$. Both periods are fixed on this substrate
while the two sizes are not: $L(\alpha)$ varies with the resources the DU
schedules, and $G(\theta)$ is what the configuration $\theta$ sets through
Eq.~(\ref{eq:cti-map}). Because the two share a base clock, the combs repeat
over a hyperperiod of $500\,\mu$s, one radio slot against four PON frames.

The handler evaluates the steady-state bound by calling the kernel function $H(\alpha, \theta)$. The returned delay bound $h_{\text{win}} = H(\alpha, \theta)$ depends on the provisioned grant size $G(\theta)$ and, by extension, the provisioning ratio $r$. The physical control point is the queue at the ONU, where each arrival waits for the next grant. Instantiating the decomposition Eq.~(\ref{eq:decomp}) from the ingress $g_{\text{in}}$ (ONU) to the egress $g_{\text{out}}$ (OLT--DU link), the end-to-end metric is
\begin{equation}
\Delta = h_{\text{win}} + \varepsilon + d^{\text{prop}}_{\text{ONU}}
+ d^{\text{proc}}_{\text{OLT}},
\label{eq:cti-bound}
\end{equation}
where $h_{\text{win}}$ is the value the kernel returns at the ONU, the span's single controlled point, $d^{\text{prop}}_{\text{ONU}} = 25\,\mu$s is the one-way propagation over the default 5\,km of fibre (the Card's $5\,\mu$s/km slope times the reach in $\rho_0$), and $d^{\text{proc}}_{\text{OLT}}$ is the processing at the OLT (FEC decode; Table~\ref{tab:card}). The term $\varepsilon$ is a fixed latency margin covering what the kernel's fluid service curve leaves out: the sub-grant quantum an arrival tail carries when it crosses a grant edge, and the RU--ONU serialization tail. Both can be a priori bounded---the serialization tail by the largest arrival unit divided by the line rate---and Section~\ref{subsec:static} sizes them against the oracle. The OLT contribution is a constant, independent of $\theta$ and of the announced regime. It plays no part in the search and enters $\Delta$ additively. 

The ONU contributes no separate transmission term. Serializing an arrival onto the upstream link takes $L(\alpha)/R$ at line rate $R$, whether or not the arrival is served across several grants, since a split shows up as waiting rather than as transmission. Nancy evaluates a whole-packet service curve, so its horizontal deviation already includes that time. The kernel therefore returns $h_{\text{win}} = d^{\text{queue}}_{\text{ONU}} + d^{\text{trans}}_{\text{ONU}}$, the partitioning freedom Section~\ref{subsec:decomp} noted.

The handler calls the kernel with the arrival unit $L(\alpha)$, the capacity $G(\theta)$, and the two periods, and the kernel returns the \emph{beat} of the two combs---how their relative phase advances from one arrival to the next---in the two forms ($h_{win}$ and $h_{edge}$) Appendix~\ref{app:comb} develops. The
forms answer different questions, and the handler uses both. The
grant-edge form $h_{\text{edge}}$ is a closed form, so it can be evaluated
independently of the kernel and used to check that the model the handler
built is the model the kernel received; Eq.~(\ref{eq:hedge}) reduces here
to the fourteen arrivals of the hyperperiod. It is rate-blind, charging an arrival its full wait to the grant edge. The
window form $h_{\text{win}}$ is the one the contract carries, because it
credits the arrival the time it spends already draining onto the line, and
that credit is the only place the line rate enters. The difference between
them is small against the grant spacing on this substrate.

Under FBA-SF a grant falls every $125\,\mu$s while arrivals come every
$35.7\,\mu$s on average, so several accumulate between grants. Under FBA-QF the
grants are four times as frequent and a single arrival instead spans several of
them. Either way bytes cross grant boundaries, which is the accumulation
Eq.~(\ref{eq:hedge}) exists to bound, and the accumulation form collapses toward the pure
wait. The same construction covers both, and in Section~\ref{sec:tsn} it
covers the 5G--TSN bridge, which shares the FBA-QF ordering with
altogether different sizes and periods.

This completes what step~3 can settle. Sweeping $\Theta_m$ and keeping what satisfies Eq.~(\ref{eq:feasible}) fixes the least $r$ at which Eq.~(\ref{eq:cti-bound}) still meets $\mathcal{D}_\alpha$ across $A$, and Eq.~(\ref{eq:cti-map}) turns that $r$ into the mapping $\alpha \mapsto \theta$. Section~\ref{sec:eval} reports the values. But it is a static answer. Each of its entries is computed as though its regime had always been in force, whereas $L(\alpha)$ changes as the DU reschedules, and $G(\theta)$ must change with it. That change is not instantaneous: the OLT applies a new grant on an actuation grid the substrate fixes, after an actuation delay that varies. What happens between the two settled points is what step~4 must resolve using the oracle. In the next Section we describe the evaluation of both step~3 and the remaining steps of the synthesis. 

\section{Framework Evaluation}
\label{sec:eval}

\subsection{Evaluation setup and the search outcome}
\label{subsec:request}

We evaluated the workflow end-to-end on the fronthaul use case as outlined in Section~\ref{sec:cti}. As the kernel we used the extended Nancy of Appendix~\ref{app:comb}, and as the oracle the ns-3 XG-PON fork. Every cost reported here is provisioning-time: nothing ran in a live network, and the closest to live operation is the joint co-simulation that closes Section~\ref{subsec:characterize}. Section~\ref{subsec:repro} quantifies the use of LLMs in the synthesis.

We started the synthesis with a request specifying the boundary and the default deployment: the default $g_{\text{in}}$, $g_{\text{out}}$, and one attached ONU. The client's declaration (step~1) then presented the fronthaul regime set $A$: the $500\,\mu$s slot pattern with its arrival sizes, the declared jitter bound, the announcement lead $\lambda_c$=4\,ms, and the minimum onset spacing $T_c = 1$\,ms (8 PON frames of 125 $\mu$s). The client's equipment class, fixed at registration in an operating network, was set to a latency requirement $\mathcal{D}_\alpha = 100\,\mu$s. The search space was the PON Card's three mechanisms (reactive DBA, FBA-SF, FBA-QF) with the per-ONU AIR as the FBA control point. The uncontrolled parameters $\Phi$ were exercised as Section~\ref{subsec:overview} defines them: every kernel evaluation was worst-cased over the relative phase of
the arrival comb (periodic process) against the service comb, the declared arrival jitter was swept across its range, and the shared clock was stressed directly (Section~\ref{subsec:dynamic}).

The declared set $A$ is continuous in the arrival size, so the handler
samples it rather than enumerating it (Section~\ref{subsec:decomp}), and
we report two samplings chosen for what each has to show. The
steady-state grid, on which we validate the kernel's worst-case delay,
takes four arrival sizes across the upper part of the declared range,
$30.6$ to $46.4$\,kB, where the grant quantization of
Appendix~\ref{app:comb} binds hardest and the computed value is therefore
most exposed; crossed with both FBA schemes and both line rates (25 and
50\,Gb/s), this gives sixteen cells. The dynamic runs of
Section~\ref{subsec:dynamic} and the cost--performance profile of
Section~\ref{subsec:characterize} extend the sampling down to $13.4$\,kB,
the bottom of the declared range, because that is where following the
traffic differs most from provisioning for its peak, and so where the
contract's efficiency is decided. The registration space $\mathcal{P}$
covered the fibre reach to 10\,km and up to four coexisting RU/ONU
tenants, and the cost--performance profile spans fifteen levels over the
full range.

The search resolved quickly. The handler drove the
kernel through the steady-state check (step~3), which settled the two FBA
schemes: FBA-QF met the requirement at every grid cell, while FBA-SF
exceeded it at every cell and was rejected as infeasible for this
equipment class. The reactive DBA failed differently. Its limiting term
was not a steady-state queue but its own control loop---observe, report,
grant---which the kernel could not evaluate. The handler therefore carried its evaluation directly to the oracle, so the mechanism passed step~3 undecided and was rejected at step~4, which measured a worst-case latency of $418$--$456\,\mu$s across transitions, breaching $\mathcal{D}_\alpha$ at every point of the profile (Section~\ref{subsec:characterize}). Note that worst-case latency reported is the maximum over the run, as is every latency we report except the trajectory of Fig.~\ref{fig:traj}, which shows the run itself. The observe-then-grant round trip sets that maximum, not the queue, which is why it barely moves with the arrival size. The residual spread is how much of a given change the report happened to carry. In the end, FBA-QF thus advanced as the sole admitted mechanism, and the synthesis completed through step~6. The following subsections report the evaluation of each step in turn.

\subsection{Soundness and tightness of the steady-state bound}
\label{subsec:static}

This subsection evaluates step~3---the handler computing the steady-state
metric $\Delta$ of Eq.~(\ref{eq:cti-bound}) through the extended
kernel---together with the first action of step~4, in which the handler
verifies each computed value against the worst-case the ns-3 oracle
measures, the check of Eq.~(\ref{eq:verify}).

The provisioning ratio $r$ comes first, since the mapping rests on it. The
fitted edge of Eq.~(\ref{eq:cti-edge}) came out nearly proportional on
this substrate: $r_{\min}(\alpha)$ ran $1.0195$--$1.0199$ across the
evaluated grid, a spread of well under a tenth of a percent, so the flat
form the Card assumed was right and only its magnitude was wrong, two
percent against the stated five. An earlier sweep had put the drift at
$1.028$--$1.044$ and suggested a load-dependent margin. Refining the sweep
showed that spread to be an artifact of the sweep resolution rather than substrate behavior. We report the corrected fit, and note the episode as the general caution: the coefficients belong to the substrate and to the resolution it was measured at, and a different substrate or a coarser evaluation grid returns different ones. What transfers is the fitting, not the numbers.

\begin{figure}[!htbp]
  \centering
  \includegraphics[width=\columnwidth]{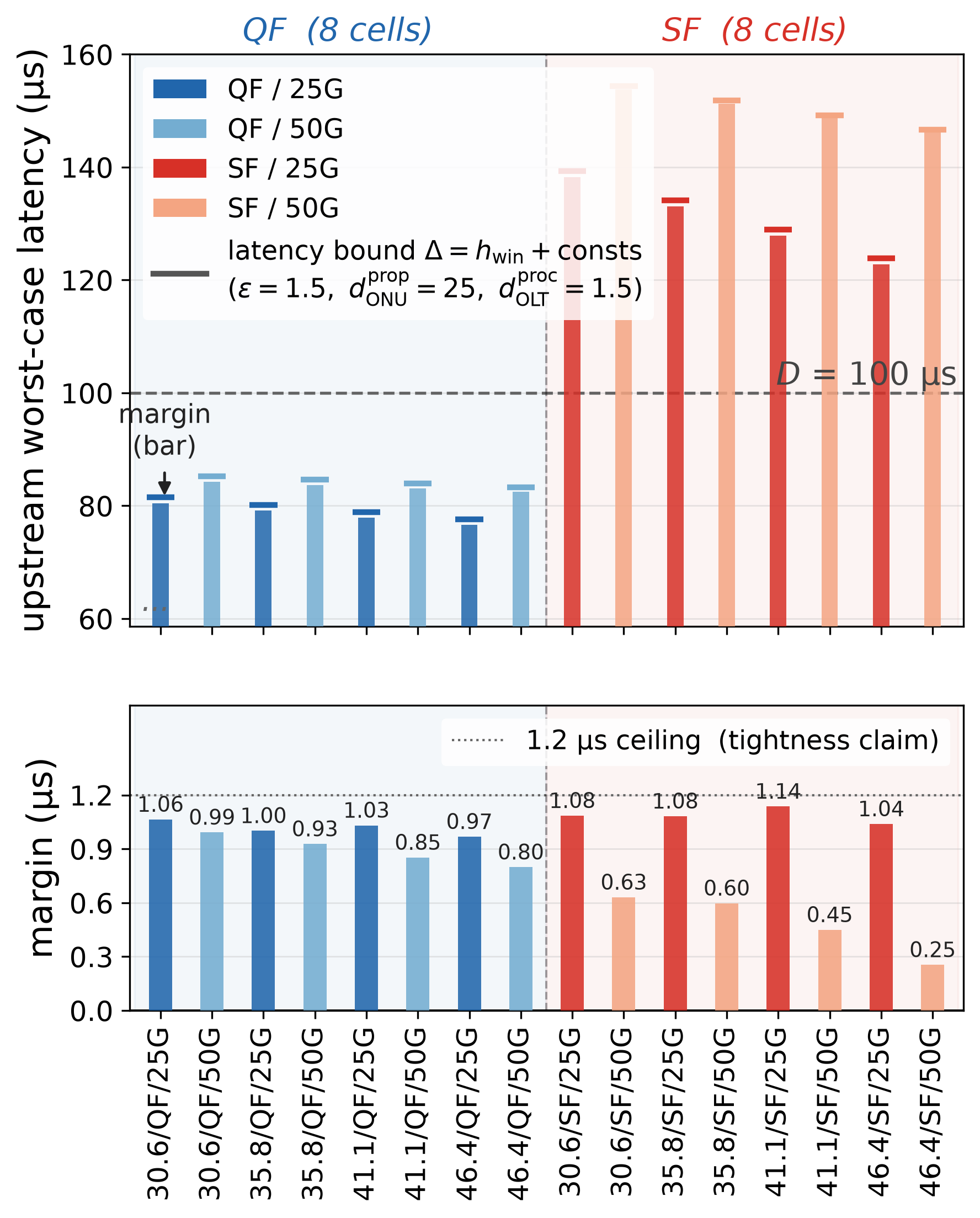}
\caption{Soundness and tightness of the kernel steady-state worst-case
delay $h_{\mathrm{win}}$ across the sixteen-cell grid. The handler adds the three delay terms the kernel does not model:
the propagation $d^{\text{prop}}_{\text{ONU}} = 25\,\mu$s, the OLT processing
$d^{\text{proc}}_{\text{OLT}} = 1.5\,\mu$s, and the fluid-model margin
$\varepsilon = 1.5\,\mu$s (Eq.~(\ref{eq:cti-bound})) to obtain the latency bound $\Delta$. (a) The latency bound $\Delta$ (caps) sits above the oracle's simulated worst-case latency (bars) in
every cell. FBA-QF meets $\mathcal{D}_\alpha = 100\,\mu$s at every cell; FBA-SF exceeds it
everywhere and is rejected at step~3, yet its computed value stays sound.
(b) The margin is strictly positive and broadly uniform across schemes,
rates, and arrival sizes.}
  \label{fig:bound}
\end{figure}

With the mapping settled, we turn to the kernel's accuracy. For each cell
the handler called the kernel for the rate-aware $h_{\text{win}} = d^{\text{queue}}_{\text{ONU}} + d^{\text{trans}}_{\text{ONU}}$ and
added the two closed-form constants of Eq.~(\ref{eq:cti-bound}),
$d^{\text{prop}}_{\text{ONU}}$ and $d^{\text{proc}}_{\text{OLT}}$, which
the kernel does not model, to obtain the end-to-end worst-case latency $\Delta$. That latency was then compared against the oracle $O$, which models the full span. Three of its delay terms were zero in this setup: $d^{\text{proc}}_{\text{ONU}}$ and $d^{\text{trans}}_{\text{OLT}}$ were configured to zero, and $d^{\text{queue}}_{\text{OLT}}$ was zero because no other traffic was present. With the margin $\varepsilon$ of Eq.~(\ref{eq:cti-bound}) set to $1.5\,\mu$s sized below, $\Delta$ sat above the oracle's worst-case in all sixteen cells, with no breaches (Fig.~\ref{fig:bound}a). The realized margin was strictly positive, spanning $0.25$--$1.14\,\mu$s across schemes, rates, and arrival sizes (Fig.~\ref{fig:bound}b): the kernel over-stated the measured worst-case by at most about $1.2\,\mu$s, so the contract did not force severe over-provisioning anywhere on the operating grid. The FBA-SF cells, although rejected at step~3, were verified all the same at step~4. A computed worst-case must be sound on infeasible candidates too, and it held.

The two kernel $h$-forms of Appendix~\ref{app:comb} carry different claims and were checked against different references. The window form $h_{\text{win}}$ is the one $\Delta$ carries, and the paragraph above describes its evaluation against the oracle, that is, against the substrate. The grant-edge form $h_{\text{edge}}$ never enters $\Delta$. It is closed form, so the handler can evaluate Eq.~(\ref{eq:hedge}) independently and compare it with what the kernel returns for the same rate-blind model; the two agreed to base-clock resolution in every cell---for FBA-SF the reference is $142.969\,\mu$s. That check validates the model the handler built against the model the kernel received, and contains no physics. The difference between the two forms is the drain credit, the time an arrival spends
already draining onto the line. It follows the grant duration $G/R$, so it
moves with the grant size as well as with the line rate, and on this
substrate it is small against the grant spacing, which is why the computed
worst-case moved only marginally between 25 and 50\,Gb/s. The line rate
also enters $h_{\text{win}}$ through serialization, which the whole-packet
curve folds in (Section~\ref{subsec:cti-comb}).

Neither kernel form covers the two residues $\varepsilon$ absorbs
(Eq.~(\ref{eq:cti-bound})). We sized its $1.5\,\mu$s from their measured
values: the sub-grant quantum at ${\le}0.25\,\mu$s and the serialization tail
at ${\le}1.1\,\mu$s. The realized margin reported above is what remains after
it. Note that $\varepsilon$ is not the fixed term $b$ of the provisioning
edge: $\varepsilon$ is a latency added to the kernel's value, $b$ a byte
quantity added to a grant, and they cover different residues.

The computed worst-case is also conditional on the declared arrival
jitter, and we measured where the condition binds. The client declared a
jitter bound of $10\,\mu$s (Table~\ref{tab:card}), and within it the
kernel's value held as stated, the measured worst-case translating by the
declared jitter. Driven beyond the declared jitter bound, to $35\,\mu$s, the
worst-case latency no longer occurred at the largest arrival size but at an
intermediate one. The fluid model assumes the worst point sits at the edge
of the declared range, so that testing the extremes suffices; at that
jitter the assumption fails, and the extremes no longer bound the
interior. The contract therefore carries the jitter bound as a validity
condition, declared once at step~1: a
client exceeding its declared jitter is outside the assume, and the
guarantee no longer applies.

\subsection{The disposition instant and its robustness}
\label{subsec:dynamic}

\begin{figure*}[!htbp]
  \centering
  \includegraphics[width=\textwidth]{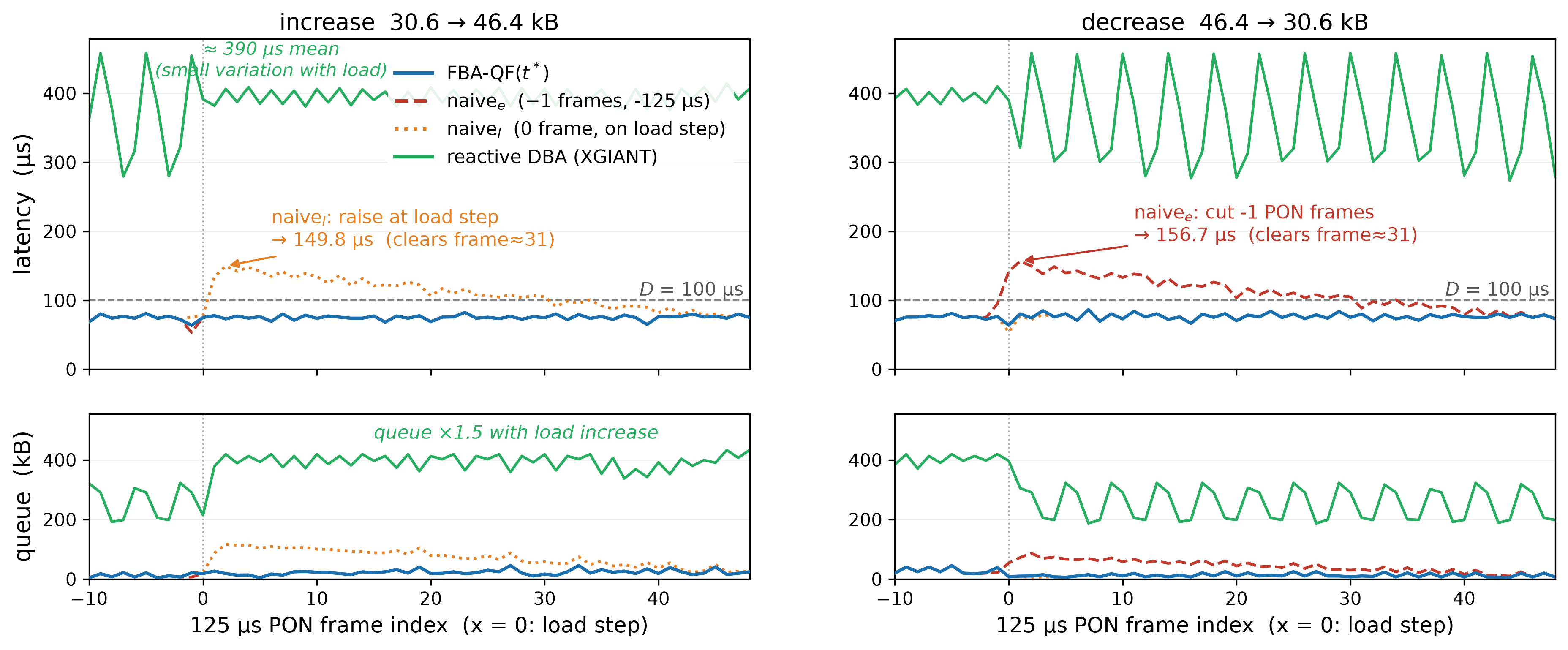}
\caption{Reconfiguration traces through the load step in both directions,
end-to-end latency and ONU queue occupancy. FBA-QF($t^{*}$) holds within $\mathcal{D}_\alpha$ each way
(peaks $86.5$ and $82.4\,\mu$s), the queue returning to the load's own
floor. The naive arms differ from it only in the disposition instant.
naive$_e$ (one PON frame early) is benign on the increase but cuts ahead
of the falling load, peaking at $156.7\,\mu$s; naive$_l$ (on load step)
mirrors it, benign on the decrease and peaking at $149.8\,\mu$s on the
increase. Both breaches clear about $31$ PON frames after the step (roughly 4x the declared spacing $T_c = 1$\,ms). The reactive DBA (XGIANT) sits on
its reaction plateau in both directions, its mean latency barely moving with the arrival size while its queue scales with it.}
  \label{fig:traj}
\end{figure*}

This subsection evaluates step~4: the handler's use of the oracle---the
ns-3 simulator with the XG-PON extensions---standing in for the live
control and data planes. The reactive DBA is the one extension to what
Section~\ref{subsec:static} covered, since it had no kernel model and was evaluated, and rejected, on the oracle. The rest of
step~4 pertains to dynamic reconfiguration. In the operating network the client
announces a new regime and the carrier re-provisions to match, and the
handler must resolve such transitions per
Appendix~\ref{app:safety}---select the side from the direction of the
change and determine the disposition instant $t^{*}$ on that side---by
running reconfiguration scenarios on the oracle at the declared spacing
$T_c = 1$\,ms. On this use case the declared set is totally ordered by the arrival size, so heavier and lighter reduce to a one-dimensional load and the two
regime change directions are a step up (increase) and a step down (decrease). The disposition instant is a
property of the transition: each direction has its own $t^{*}$. The handler sampled the extremes transitions first, lightest-to-heaviest and heaviest-to-lightest, then intermediate pairs; the measured instant was the same across the samples in each direction, so the mapping collapsed to one $t^*$ (Fig.~\ref{fig:traj}).

The synthesized solution is FBA-QF applied at the disposition instant the
handler determined, labelled FBA-QF($t^{*}$) in the figures. Each
direction carried its own instant: one PON frame ahead of the load step (onset)
for the increase, and at the step for the decrease. The handler reached these
through a sequence of oracle test scenarios, failed candidates included, and the figures show the resolved endpoint, which is why those traces look smooth as if no transition happens. It held $\mathcal{D}_\alpha$ both ways, with transient peaks of
$86.5$ and $82.4\,\mu$s and the queue returning to the load's own floor.
These are the \emph{return} and \emph{bounded-transient} conditions of
Eq.~(\ref{eq:safety}), met on the extreme transitions of the declared set
in both directions. 

To isolate what the disposition instant contributes, we drove three alternatives
through the same load steps. These are comparison schemes we scripted, not
outputs of the handler. The first two isolate the disposition instant with
a fixed apply in both directions: naive$_e$ applied one PON frame before
the step (as $t^{*}$ in load increase - but fixed always), naive$_l$ at the step (as $t^{*}$ in load decrease - but fixed always), and each differed from the 
synthesized solution only in the disposition instant. On its safe side
each of the two schemes landed inside the safe set and reproduced the synthesized trace
and its peak. On the wrong side each breached. The naive$_e$ decrease cut the grant one frame before the load fell and peaked at $156.7\,\mu$s, and the
naive$_l$ increase left the old, smaller grant facing the already-heavier
load for one frame and peaked at $149.8\,\mu$s. Both were transient violations of the \emph{bounded-transient} condition, and they
cleared about 31 PON frames after the step, roughly $3.9$\,ms. The declared
spacing $T_c$ is $1$\,ms. So these schemes fail the \emph{return} condition as well: the queue is still draining when the next announced change may arrive. Eq.~(\ref{eq:safety}) 
fails on both counts, the \emph{bounded transient} and the \emph{return}, and the latter is the one that compounds.

The third scheme is the reactive DBA (XGIANT), polled at $0.5$\,ms---faster than practical deployments run it, so the mechanism is reported at its most favourable setting. The same runs are its step-4 rejection evidence and its baseline here. It sat on its reaction plateau in both directions: its mean latency barely moved with the load while its queue scaled with it, from $253$ to $403$\,kB. Bytes that arrive and need a larger grant wait for the loop to notice them, so the wait is set by the loop's period rather than by how many bytes are waiting. The plateau is not a steady-state property. At a fixed load served by a grant already sized for it, the loop would be unremarkable. What it cannot absorb is a change. At every transition---including the first, from an empty queue to any load---the queue builds while the report and the grant travel, and it then stays there: the new grant is sized to serve the arrival rate, not to drain what accumulated while it was being computed. The queue settles at a new equilibrium carrying the transition's backlog, and the latency sits on the plateau with it. Where FBA-QF($t^{*}$) returns to the load's own floor after each change, the reactive loop does not return at all. Its worst case ran higher and reached $448\,\mu$s at the lighter regime, with six times the standard deviation, so it is also least predictable where it is least loaded. This is the \emph{return} condition of Eq.~(\ref{eq:safety}) failing permanently rather than transiently, and it is why the mechanism died at the oracle rather than at the kernel---the limit is the loop, not a queue.

The chosen disposition instant proved robust to imprecise actuation, and
the handler measured that robustness on the oracle rather than assuming
it. Actuation on this substrate is frame-quantized---the FBA applies a
configuration at the boundary of a $125\,\mu$s PON frame---so $t^{*}$ is
enacted on that frame, while the actuation delay within it varies at
sub-frame scale. The measured safe sets spanned multiple frames, which is
the margin Appendix~\ref{app:safety} requires between the safe set and the
$\Phi$-scale imprecision. The shared clock was stressed directly: the
handler injected a clock-rate drift of $\pm1000$\,ppm, roughly $200\times$
a realistic PTP holdover of 4.6\,ppm, and the computed worst-case latency held within $\mathcal{D}_\alpha$ at every load but one. The one excursion, a slow beat at the highest 
load, scaled linearly with the drift and shrank to ${\sim}0.5\,\mu$s at
the realistic holdover, so it marks the drift axis of the contract's
validity rather than a failure at any plausible clock. As a second fault
test, the announcement itself was driven out of contract: declarations
delayed past the committed lead $\lambda_c$ produced a divergence that
grew with the fault, so a violated commitment shows up as a violated
guarantee, and the fault is attributable to the announcement rather than
to the carrier. Finally, the decrease-direction $t^{*}$ was checked down
to the stability floor. Appendix~\ref{app:safety} leaves one question open
per substrate: whether stable operating points exist that admit no safe
disposition instant---whether $\mathrm{SAFE}(T_c)$ is a proper
subset of $\mathrm{STABLE}$. On this substrate, at the
declared spacing, the answer is no. At every provisioning level down to
the floor, the frame boundary one ahead of the onset remained a safe disposition instant
at the mapped configuration, with no raise to a $\theta^{+}$ required, so
every stable operating point is safely reconfigurable and the two sets
coincide.

\subsection{Characterization: reach, tenancy, and cost--performance}
\label{subsec:characterize}

At step~5 the handler used the oracle to characterize the admitted
contract across the registration space, and each axis returned a compact
answer. First, reach. Two slopes fold together: each kilometre of fibre
adds $5\,\mu$s of propagation to the (data-plane) latency bound, while the
equalization delay---the delay the OLT assigns each ONU so that ONUs at
different distances align on one virtual distance---advances the effective apply by the same slope, shortening the required lead
as reach grows. At zero reach that lead is $135\,\mu$s: twice the propagation to the farthest ONU at the $10$-km limit, plus $35\,\mu$s of header processing. The handler characterized the envelope to the 10\,km substrate
limit. A link is admitted when its latency bound at that distance meets the
registered deadline $\mathcal{D}_\alpha$, which for the $100\,\mu$s class binds the admissible
reach close to 6.5\,km. Second, tenancy. Coexisting RU/ONU
tenants were added at both line rates, each held at or below $85\%$ frame
occupancy. Across the feasible cells the guarantee moved by at most
$0.04\,\mu$s, and the one infeasible cell---four tenants at
25\,Gb/s---failed on capacity packing, with no loss of isolation between the flows. Together the two axes give the
validity envelope $\mathcal{P}_m$ the registration interface consumes.

\begin{figure}[!htbp]
  \centering
  \includegraphics[width=\columnwidth]{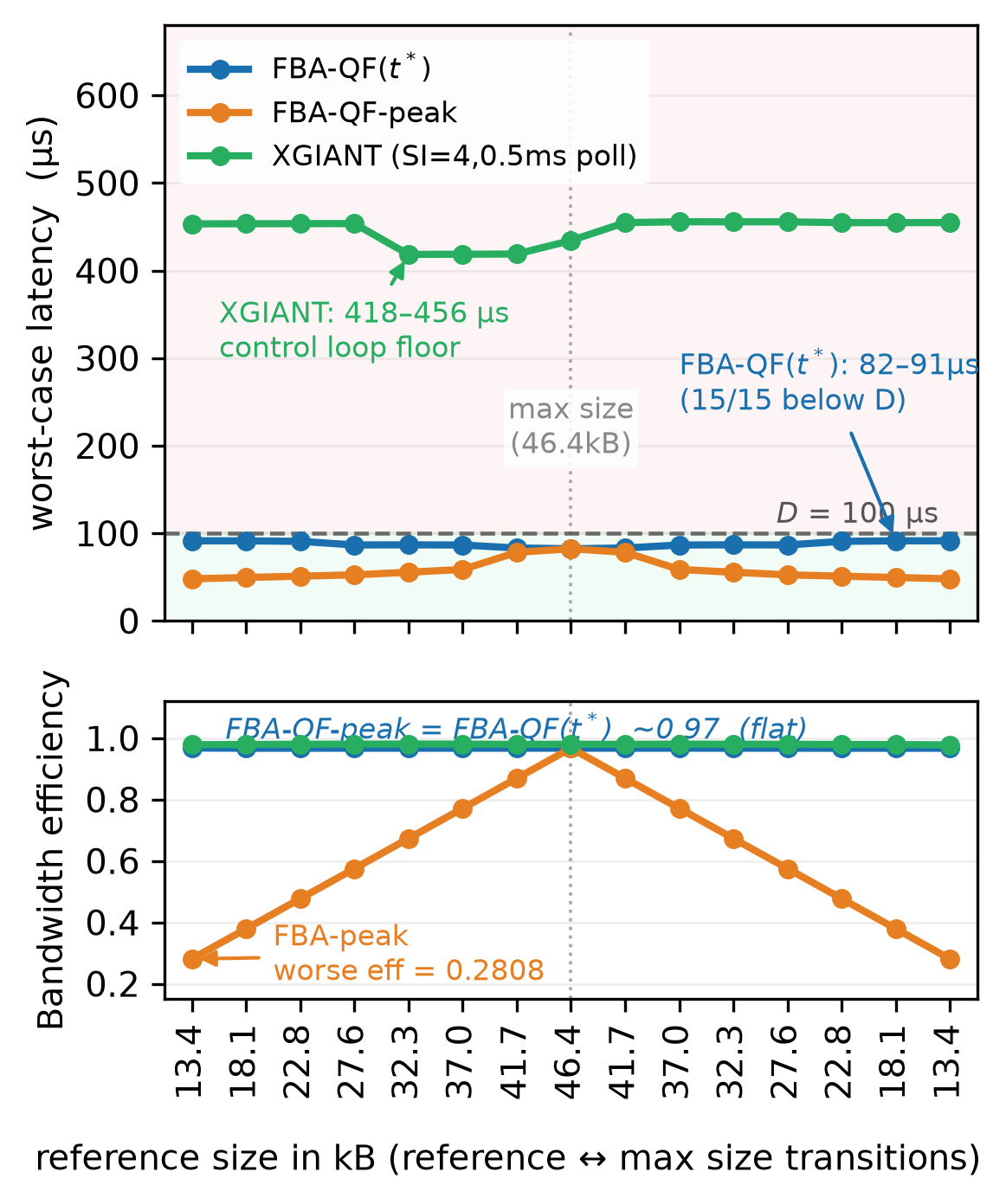}
\caption{Latency versus efficiency over the fifteen-level load profiles (repeated transitions from reference to max size). FBA-QF($t^{*}$) is the only scheme in the meets-deadline, high-efficiency corner ($91\,\mu$s, $0.97$). The FBA-QF-peak meets $\mathcal{D}_\alpha$ but wastes bandwidth at low load ($0.28$); the reactive DBA (XGIANT) is the most efficient ($0.98$) yet breaches $\mathcal{D}_\alpha$ at every level, its lowest worst-case above $418\,\mu$s at a $0.5$\,ms poll---the irreducible observe-then-grant round trip.}
  \label{fig:eff}
\end{figure}

Third, cost--performance, where we evaluate the efficiency defined as $x=\ell(\alpha)/p(\theta)$, the reciprocal of the provisioning ratio $r$ of Eq.~(\ref{eq:cti-ratio}) in this use case. Note that stability requires $x \le1$. Here the comparison is with the alternative
the framework replaces. The peak-provisioned FBA (FBA-QF-peak) is what a static contract
over the same declared set $A$ would amount to: one configuration, sized for the heaviest regime ($L = 46.4$\,kB), held at all times. It is also within the framework's reach: steps~3--4 would seek a single $\theta$
admitting every $\alpha \in A$ instead of a mapping, with no $t^{*}$ to place.
So it is not a straw baseline but the honest static answer to the same problem. Across the fifteen-level load profile
(Fig.~\ref{fig:eff}) FBA-QF-peak met the deadline, but its efficiency fell toward the light end, to 0.28,
paying for the peak at even light load. FBA-QF($t^{*}$) met the same deadline
at $0.97$ efficiency, a factor of $3.5$, because it follows the declared regime instead of bounding it. Note that efficiency is measured on the deployed grant, which rounds up from the fitted edge to the substrate's allocation quantum, explaining the difference of ${\sim}1\%$ with respect to the inverse of the provisioning ratio $r$.
We also plot the reactive DBA (XGIANT), which reached $0.98$ efficiency but
breached $\mathcal{D}_\alpha$ at every level, its worst-case ranging $418$--$456\,\mu$s---the
$0.5$\,ms poll of Section~\ref{subsec:dynamic}, the mechanism at its best. The ramp runs in both directions, and the two proactive schemes retrace their outbound curve, since a grant follows the announced regime and nothing else. The reactive DBA does not. Its grant follows the queue with the loop's lag, so on the way up it sits behind a rising load and on the way down ahead of a falling one, and the same level is not the same operating point in the two legs, explaining thus the asymmetry observed in Fig.~\ref{fig:eff}. Concluding, the three compared arms separate cleanly: the static provisioning FBA-QF-peak holds the deadline and wastes the capacity, reactive DBA scheduling recovers the capacity and loses the deadline, and only the dynamic contract FBA-QF($t^{*}$) holds
both. That is the efficiency the interface exists to buy, and the
disposition instant is what makes it safe to take.

With a single admitted mechanism the selection of step~6 was immediate,
and the workflow emitted its outputs: the dynamic contract $\mathcal{C}$, the two 
adapters, and the registration interface. As the final test---the closest
this evaluation comes to a live network, with the oracle standing for
it---we exercised the emitted artifacts in a joint co-simulation. The
synthesized 5G adapter sat above a 5G traffic generator, its announced
configurations were transferred to the OLT and applied at $t^{*}$, and a
single ns-3 run exercised the whole contract: announcement, timed
configuration, transition, and settled guarantee.


\subsection{Reproducibility and the division of labor}
\label{subsec:repro}

In our tests a single LLM runs steps~2 through~5 over the PON Card,
constructing the network model, writing the scenarios, and calling the
kernel and the oracle as tools. The question is whether that role is
dependable, and whether the numeric authority we withhold from the LLM could
safely be returned to it. We measured both with fresh LLM draws on the frozen
Card: $K = 6$ independent synthesis runs, three each on two LLM families
(Claude Opus~4.8 and Gemini~3.1~Pro, drawn in June 2026), scored per property in
Table~\ref{tab:brain}, with eighteen draws across the three PON conditions; Table~\ref{tab:brain} also carries six draws on the 5G Card (Section~\ref{sec:tsn}).

The constructive role reproduced. All six base draws completed the
workflow---network model constructed, kernel driven, disposition instant
reasoned, mechanisms ranked, with the reactive DBA excluded from the
kernel path and rejected on the oracle's measurement---and all six
surfaced the Card's defects: the overhead entry
(Section~\ref{subsec:card}), an unquantified reference to how a fragment
spills across a grant edge, and an ambiguous reach basis. A surfaced defect is reported and, where it does
not block the workflow, carried forward. Six draws per condition across two
families is a reproducibility measurement and a designed ablation, not a
statistical safety proof. The Card is also not a sanitized input: carrying
real defects the draws must catch is part of the framework.

The numeric role did not reproduce, and the two ablation conditions fail in
opposite directions. We asked the LLM to produce the worst-case delay directly,
bypassing the kernel, and five of six draws were optimistically unsound. They treated the unquantified fragmentation spill as zero, and their values, assembled into $\Delta$ with the span's constants, came to $\sim71.5\,\mu$s against the oracle's measured $80.1\,\mu$s. The
sixth declined to compute what the Card did not quantify and stayed sound
at ${\sim}102\,\mu$s, pessimistic and exceeding the QoS deadline. Optimism is the dangerous direction: a value below
the substrate's own behavior promises what the carrier cannot deliver, and
no downstream check catches it, because that value is what the checking
would be against.

We then closed the loop the draws themselves had opened. The missing
quantity was the grant-fill quantization---how a per-tooth grant is sized
by ceiling word-division, so that an arrival spanning two grants leaves a
residual---which is the mechanism separating $h_{\text{edge}}$ from
$h_{\text{win}}$ in Appendix~\ref{app:comb}, and it had no Card entry. We
added one field carrying it and ran six fresh draws on the same two
families, with the query held verbatim. All six were then sound: the
emitted values after adding the delay constants $\Delta$ moved up by $18$ to $31\,\mu$s, from the breach at
$71.5\,\mu$s into the range $89.1$--$102.7\,\mu$s, every one of them above
the measured worst-case. A Card field can carry mechanics into the
reasoning, and doing so repaired the direction of the error. However, it did not repair the magnitude, and the error spread is where the division of
labor shows. The kernel's value with the delay constants added comes to $84.2\,\mu$s for that cell, loose by at most $1.2\,\mu$s against the oracle's measurement (Section~\ref{subsec:static}). The tightest draw reached $89.1\,\mu$s. The
loosest reached $102.7\,\mu$s, which exceeds the $100\,\mu$s deadline and
so declares infeasible a configuration the kernel and the oracle both
admit. That is a false negative rather than a broken guarantee, an error
in the safe direction, but a synthesis that made it would reject the
contract it was asked to produce. Soundness is what a Card field can
supply. Tightness remained the kernel's. The synthesis takes neither the LLM's soundness nor its tightness on trust: the kernel computes the number, the oracle admits it, and no probabilistic output enters the contract. 

One analytic result was never reproduced. The kernel offers two forms for the same queue: $h_{edge}$ charges an arrival its full wait to the edge of the grant that clears it, $h_{win}$ credits the time the arrival spends already draining onto the line (Appendix~\ref{app:comb}). The second is tighter and moves where the worst case sits across the load range. In twelve of twelve draws across both conditions, no draw derived that second form, nor the trend that follows from it. What the draws could not do is derive. Computing the refinement is the kernel's job, and the handler's job is to call the kernel.

The computation time, finally, is
provisioning-time: the kernel priced a cell in $52$ to $183$ seconds, the
oracle runs parallelize, and the whole synthesis re-runs cheaply against a
corrected Card.

\begin{table*}[!htbp]
\centering
\caption{Brain test on the frozen Domain Card. Six independent draws ---
three fresh contexts on each of two LLM families (Claude Opus~4.8,
Gemini~3.1~Pro), memory off, isolated directories, with every
load-bearing number traced to a named Card field. Each row is one
property of the synthesis, scored over the draws.}
\label{tab:brain}
\small
\begin{tabular}{@{}p{0.19\textwidth} p{0.09\textwidth} p{0.64\textwidth}@{}}
\toprule
\textbf{Property} & \textbf{Score} & \textbf{What the draws did} \\
\midrule
\multicolumn{3}{@{}l}{\textit{Framework: PON Card frozen - the LLM constructs network model and asks the kernel for worst-case delay calculation}}\\
Network model and structure & $6/6$ & The same two-comb model, the queue at the ONU, and the same additive bound structure, each independently re-derived --- including the commensurability of the $500\,\mu$s radio slot with four $125\,\mu$s PON frames, which the Card does not state. Every draw reached the same configuration and the same frame for the disposition instant. \\
Provenance honesty & $6/6$ & The deadline was flagged as external to the substrate, with a transparently labelled candidate value. No draw asserted it as a substrate fact or invented a source. \\
Card defects surfaced & $6/6$ & All three defects found: the overhead-optimistic FEC ratio, an unquantified fragmentation-spill reference, and an ambiguous reach basis in the reaction-time bound. None is expressible as a schema check. \\
Mechanism ranking & $6/6$ & The reactive DBA ruled out on its reaction time; FBA-SF infeasible for the $100\,\mu$s equipment class and feasible for a looser one; FBA-QF selected --- each draw judging against its own labelled candidate deadline. \\
\addlinespace
\multicolumn{3}{@{}l}{\textit{Ablation: the kernel is bypassed, the LLM asked for worst-case delay calculation}}\\
Base Card & unsound $5/6$ & Five draws treated the unquantified grant-fill spill as zero and emitted ${\sim}71.5\,\mu$s against the measured $80.1\,\mu$s. The sixth flagged the Card's gap, declined to fold the spill to zero, and stayed sound at ${\sim}102\,\mu$s. Optimistic error is the dangerous direction: nothing downstream catches a value that is itself the reference. \\
\addlinespace
Amended Card & sound $6/6$ & One field added, carrying the grant-fill quantization the draws had flagged---the mechanism separating $h_{\text{edge}}$ from $h_{\text{win}}$. Query held verbatim. Emitted values moved up $18$--$31\,\mu$s into $89.1$--$102.7\,\mu$s, all above the measurement. Direction repaired. \\
\addlinespace
Tightness after repair & $84.2\,\mu$s vs $89.1$--$102.7\,\mu$s & The kernel's bound against the amended draws' spread. The loosest exceeds $\mathcal{D}_\alpha=100\,\mu$s and rejects a configuration that holds---a false negative, safe in direction but fatal to the synthesis. Soundness is what a Card field supplies; tightness is the kernel's. \\
\addlinespace
Window refinement & $0/12$ & Neither the refinement of Appendix~\ref{app:comb} nor its worst-at-low-load trend was derived, in the base or amended draws: all twelve placed the worst-case at the highest load, where it is at the lowest. Derivation-limited, and the one capability the architecture does not ask for. \\

\addlinespace
\midrule
\multicolumn{3}{@{}l}{\textit{Framework: 5G--TSN bridge brain test - Section~\ref{sec:tsn}}}\\
Apply framework to a second substrate & $6/6$ & Three draws on each LLM family re-derived the service comb from raw slot pattern, emitted the exact per-cell waits, and ruled the over-rate cells infeasible. The rate-edge cells passed throughout. The Opus draws named the zero-slack case, splitting $2$--$1$ on the declared completion convention; the Gemini draws applied a stability rule they flagged as external framing. \\
\bottomrule
\end{tabular}
\end{table*}

\section{Synthesizing TSN over 5G}
\label{sec:tsn}

Sections~\ref{sec:cti} and~\ref{sec:eval} used the framework to synthesize
the CTI: a dynamic interface carrying 5G fronthaul over a TDM-PON, with
the PON as the carrier. The framework is not specific to that substrate.
To show its generality we apply it unchanged to a different carrier: a 5G system transporting a Time-Sensitive Networking (TSN) flow,
acting as the 3GPP logical bridge between two TSN segments~\cite{5gtsn}.
This boundary is standardized too. 3GPP specifies the 5G system as a
transparent TSN bridge that reports a bridge-delay managed object per port pair
and traffic class, which the TSN control plane consumes when it admits a
flow. As with CTI, we target the load-bearing part: what the client
declares, and what delay the carrier can then report.

The roles invert: the 5G system is now the carrier, and it is more than the RAN that acted as client before: the 5G--TSN bridge runs from the UE, through the RAN
(RU--DU--CU), to the user-plane function (UPF), whose network-side TSN
translator hands Ethernet frames to the next TSN segment. The same
construction and the same extended kernel carry over. Of the bridge's
delay blocks we model the dominant scheduled queue---the radio uplink
stage where the TSN payload waits for its allocation---which is the
bottleneck reduction of Section~\ref{subsec:decomp}.

The interface this bridge standardizes is of the static class, and the reason is in its control plane. We
assume time-aware shaping (TAS), for which control is fully centralized:
the bridges, the 5G system among them, report their capabilities; the TSN controller computes offline the centralized network configuration (CNC) that includes the paths and gate
schedules and installs them; and each flow is admitted once, its
requirements mapped into the 5G system at establishment. The runtime then
executes and enforces the installed schedule. A traffic changeover arrives
as a new registration, recomputed and reinstalled, not as an announced
transition the bridge follows. The schedule is recomputed rather than adjusted---gate-schedule synthesis is a
global problem and the usual formulations solve it from scratch~\cite{tsn-ilp}---
so a hitless changeover is not what this control plane is built for. In the framework's terms the client
announces no change, so the contract is static: synthesis still covers the
whole declared set $A$, but there are no transitions to evaluate, so synthesis 
step~4 is bypassed and registration fixes the one regime that stays in
force (Section~\ref{subsec:outputs}). Step~3 is therefore the load-bearing
synthesis step, and we run the workflow focusing on it: the handler ingests the 5G
Card, constructs the network model, and calls the kernel, and the emitted
guarantee is exactly the bridge-delay managed object the CNC consumes at
admission. The oracle seat, which a static contract still uses for
verifying the computed value and for the step-5 characterization, is not
filled---building a faithful 5G packet-level substrate is beyond this
paper's scope---so the kernel serves as the numerical reference against the handler's closed form, a check of the model rather than of the substrate, and the reproducibility draws of
Section~\ref{subsec:repro} were repeated on this Card
(Table~\ref{tab:brain}).

\subsection{The 5G Card}
\label{subsec:tsn-card}

The carrier is a 5G system in an industrial deterministic configuration.
We build the 5G Domain Card from the 3GPP numerology for this
scenario~\cite{tdd10}: $\mu=1$ ($30$-kHz subcarrier spacing), a $0.5$-ms
slot, and the TDD-10 frame structure, which yields exactly one active uplink
allocation per millisecond, so $T_{\text{ser}} = 1$\,ms. The allocation is a
contiguous block of uplink symbols, which 3GPP calls a \emph{window}; we treat it as one \emph{grant}, the 5G analogue of the PON's grant. Its internal structure does not enter the model, because the requirement is on the burst and not on the symbols within it. The Card declares one mechanism, $M = \{\,\text{Configured Grant}\,\}$: a
3GPP pre-allocation of the service comb, one grant per period. Its
parameter space $\Theta_m$ is that pre-allocation, which sets $G(\theta)$,
the bytes a grant carries, so the provisioned rate is $p(\theta) =
G(\theta)/T_{\text{ser}}$. For strict reliability the capacity is a
transport block sent at a robust modulation and coding scheme ($64$-QAM at
code rate ${\approx}0.35$), and we hold $\theta$ at that operating point
throughout, $G(\theta) = 277$ Bytes per grant. A static contract pairs
one regime with one configuration at registration, so what synthesis must
establish is which regimes in $A$ this configuration admits, rather than a
mapping across the set. The registration context $\rho_0$ fixes the span
from the UE to the network-side translator.

The client's declaration is the regime set $A$. Each regime $\alpha \in A$
is one \emph{payload burst}, a discrete block of industrial data of size
$L(\alpha)$ bytes, arriving once per transfer interval $T_{\text{arr}}
(\alpha) = T_{\text{GCL}}$, so the offered load is $\ell(\alpha) =
L(\alpha)/T_{\text{arr}}(\alpha)$. Unlike the PON use case, where the
arrival period was fixed and only the size varied, here both vary across
$A$: the 3GPP deterministic classes span transfer intervals of one to ten
milliseconds and payloads from a few hundred bytes
upward~\cite{cps-req}. Cells A--F of Table~\ref{tab:tsn} sample those
classes, and each carries the requirement $\mathcal{D}_\alpha$ its class
states. Rows G onward leave the declared classes deliberately, walking the
interval down to a third of a millisecond to locate where the closed form
stops being exact (Section~\ref{subsec:tsn-compose}). Because the contract
is static, the client announces no onset and commits no lead: what
registration fixes is one $\alpha \in A$ together with the configuration
synthesis paired with it. As in the PON Card, the mechanism exposes one
continuous control point. What the capacity
includes (coding, overhead), how the allocation is laid out in time, and
the control plane that installs it all differ, and the Card's remaining
entries carry those differences. The handler ingests the Card with the
same provenance tagging.

\subsection{The 5G--TSN bridge-delay bound}
\label{subsec:tsn-comb}

The parameters are far from the PON ones---a single payload burst rather than a stream of radio arrivals, byte sizes and periods an order of magnitude apart, and an arrival period that varies where the PON's was fixed---but the structure is the one Section~\ref{subsec:cti-comb} already built: a periodic arrival served by a periodic comb, with the same beat producing the hard queueing term. One burst is served across several faster grants ($T_{\text{ser}} < T_{\text{arr}}(\alpha)$), which is the FBA-QF ordering, and the construction spans it without change.

The decomposition Eq.~(\ref{eq:decomp}) instantiates over the 5G--TSN
bridge just as over the PON substrate: the handler assembles $\Delta$ from
the closed-form terms the Card and $\rho_0$ fix and the hard term the
kernel returns. Stability requires the provisioned rate to exceed the
offered load, $\ell(\alpha) < p(\theta)$. Beyond that, because there is
one burst per interval and---in the industrial classes---the drain
completes before the next burst arrives, the kernel call
Eq.~(\ref{eq:kernelcall}) reduces from the general accumulation form
Eq.~(\ref{eq:hedge}) to a single drain:
\begin{equation}
H(\alpha,\theta) = h
= \Big\lceil \frac{L(\alpha)}{G(\theta)} \Big\rceil\,T_{\text{ser}},
\qquad
T_{\text{arr}}(\alpha) \;\ge\; h ,
\label{eq:tsn-h}
\end{equation}
the integer number of grants a burst waits, counted from an arrival that
just misses a grant. The side condition is what makes it a single drain:
no burst may still be draining when the next one arrives, and
Section~\ref{subsec:tsn-compose} develops what happens where it fails.
This is the grant-edge form of Appendix~\ref{app:comb} in its
one-burst-per-interval case, so it is rate-blind: it charges the burst to
the edge of the final grant and never mentions the radio rate, which is
where the drain credit would enter. The kernel needs no modification, and
its evaluation makes no assumption about which period is faster. Across
the industrial classes the closed form equals the value $H(\alpha,\theta)$
returns in every feasible cell, gap zero, including the two critical cells
where the offered load meets the provisioned rate (cells E and F,
Table~\ref{tab:tsn}).

That agreement is on the queueing term alone---the $d^{\text{queue}}$
block of Eq.~(\ref{eq:decomp}), which for this scheduled mechanism takes
the form of the scheduling wait $h$---and the scope matters.
Eq.~(\ref{eq:tsn-h}) is the exact number of grants a burst waits: under
its side condition there is nothing pending behind it, the accumulation
form has no slack to drop, and the kernel confirms it cell for cell. Two
things it does not include. The rate-aware refinement, since the form is
the grant-edge one and the credit for a burst already draining onto the
radio would need the substrate's own measurement. And how the $L(\alpha)$
bytes map onto the transport blocks that fill each grant: if a burst tail
spills past a grant period the true wait rises by up to one grant, the
same effect that, on the PON, made the fluid form lose tightness when an
arrival tail crossed a grant edge (Section~\ref{subsec:static}). There,
the packet-level oracle measured that residual and the margin
$\varepsilon$ covered it. Here there is no such oracle, so we claim
exactness for the scheduling term and no further. Both residuals are
bounded by construction rather than measured, and they are not zero, and
we do not claim they are.

\subsection{Composition and the transferred configuration}
\label{subsec:tsn-compose}

The standardized interface is static, and so is the contract we synthesize
for it. A changeover is a re-registration---the CNC recomputes the
schedule and reinstalls it---and industrial practice makes these
infrequent and not hitless. The framework would synthesize a dynamic interface for this boundary from the same machinery. A configured-grant change is the 5G analogue of the PON grant
change. The two-window criterion of Eq.~(\ref{eq:safety}) applies to it
unchanged, with the direction selecting the safe side as derived there.
What is missing is not theory but an evaluator: the disposition instant
$t^{*}$ has no algebraic form and would need this substrate's own step-4
measurement, which is why a packet-level 5G oracle is the natural next
seat to fill. What the bridge consumes today is the static part: the TSN
centralized configuration programs the gate schedules in advance over a
shared gPTP time reference, taking as the bridge's reported delay exactly
the guarantee synthesized here.

The exactness of Eq.~(\ref{eq:tsn-h}) has a limit, and the limit is
closed-form rather than intuitive. Write $x = \ell(\alpha)/p(\theta) =
L(\alpha)T_{\text{ser}}/(G(\theta)T_{\text{arr}}(\alpha))$ for the utilization---the reciprocal of the provisioning ratio $r$ of Eq.~(\ref{eq:cti-ratio}), and the same quantity Section~\ref{subsec:characterize} reported as bandwidth efficiency. Stability requires $x \le 1$. The side condition of
Eq.~(\ref{eq:tsn-h}) is stricter than stability, and it reads differently
on either side of the grant size. For a payload at least the grant size
($L \ge G$), the drain must complete before the next burst arrives,
$T_{\text{arr}} \ge \lceil L/G \rceil\,T_{\text{ser}}$. For a smaller
payload ($L < G$), one grant must absorb everything the interval
delivers, $T_{\text{arr}} \ge T_{\text{ser}}/(\lceil G/L \rceil - 1)$.
Both say the same thing---no burst waits behind another---and both bind
through $T_{\text{arr}}$, which is why the interval and not only the
payload decides where the form holds. 

The kernel's sweep confirms the
limit and its character (Table~\ref{tab:tsn}). At the limit the form is
still exact (cell G). Just below it the worst-case leaves the first drain
and lands a few arrivals in, at index $n^{*} = 3$, where $n^{*}$ is the
arrival within the hyperperiod that attains the maximum in
Eq.~(\ref{eq:hedge}); the gap is $1\mu$s (cell H). Approaching the stability floor the worst-case leaves the first arrival altogether, landing anywhere in the hyperperiod ($n^{*}$ of 31 and 6), while the gap widened to 165 and 260\,$\mu$s (rows R1 and R2). The limit is
reachable at ordinary parameters, not only in a corner: a $500$-B burst on
a $1.9$-ms interval, inside the industrial payload range, is unsound by
$400\,\mu$s (row X), while the same burst at a $2.0$-ms interval sits at
the limit and is exact (row W). A
$100\,\mu$s of interval separate row W from X, and $400\,\mu$s of
soundness.
Cells E and F sit exactly at the rate edge and stay exact because their
grant count is an integer, leaving no room between the stability floor
and the limit. 

Row X is worth reading closely, because the numbers look benign: at 
$L=500$\,B and $T_{\text{arr}}=1.9$\,ms the utilization is $x=0.95$, so
the system is stable, and the interval is nearly twice the grant period.
What fails is the side condition of Eq.~(\ref{eq:tsn-h}). The burst needs
$\lceil 500/277 \rceil = 2$ whole grants, so $h=2$\,ms, $100\,\mu$s
more than the interval. Quantization is what bites: the burst
needs $1.8$ grants of capacity but must wait for two, and the unused
remainder of the second carries forward. Each burst leaves a little
behind, the residue accumulates, and the worst-case moves off the first
burst to $n^*=4$. The single-drain form assumes nothing is pending, and
is then $400\,\mu$s optimistic. Within the industrial classes the scheduling term is exact.
Below the limit the kernel's evaluation of the general form
Eq.~(\ref{eq:hedge}) is what the handler must use, and what carries over
is the construction---the computed value, the configuration (untimed here,
since this class has no disposition instant), and the validity
limit---not a particular number.

Two carriers, PON and 5G, with quite different arrival and service
parameters, are handled by the same workflow, the same reasoning modules,
and the same kernel, and yield contracts of the same form, one per
interface class: the PON carries the dynamic contract, the 5G--TSN bridge
the static, registration-consumed one. Both use cases reduce to the
per-hop primitive of deterministic domains---a single scheduled queue, a
periodic arrival, a periodic service. That the second use case was
synthesized rather than adapted is measured, not asserted: six fresh draws
on the frozen 5G Card, three per LLM family, re-derived the service comb
from the raw slot pattern, emitted the exact per-cell waits, ruled the
over-rate cell (D) infeasible, and flagged the deadline as the one input
external to the substrate (Table~\ref{tab:brain},
Section~\ref{subsec:repro}). A domain enters as a Card, and the
construction follows.
\begin{table}[t]
\centering
\caption{The scheduling term $h$: the closed form of Eq.~(\ref{eq:tsn-h}) against the value the kernel returns, at $G(\theta)=277$\,B per $1$-ms grant. Every row probes the side condition $T_{\text{GCL}} \ge h$: exact above it, exact and critical on it, optimistic below. A negative gap means the closed form falls below the kernel's; $n^{*}$ is the arrival attaining the worst case in Eq.~(\ref{eq:hedge}).}
\label{tab:tsn}
\small
\setlength{\tabcolsep}{4pt}
\begin{tabular}{@{}c r r r r l@{}}
\toprule
Cell & $L$ (B) & $T_{\text{GCL}}$ (ms) & $h$ (ms) & gap ($\mu$s) & status \\
\midrule
A  & $200$  & $4$     & $1$  & $0$    & feasible \\
B  & $500$  & $4$     & $2$  & $0$    & feasible \\
C  & $1000$ & $10$    & $4$  & $0$    & feasible \\
D  & $1200$ & $4$     & ---  & ---    & infeasible \\
E  & $1108$ & $4$     & $4$  & $0$    & critical, exact \\
F  & $2770$ & $10$    & $10$ & $0$    & critical, exact \\
\addlinespace
\multicolumn{6}{@{}l}{\itshape crossing the limit at a short interval ($L = 78$--$80$\,B):}\\
G  & $78$   & $0.334$ & $1$  & $0$    & at the limit, exact \\
H  & $78$   & $0.333$ & $1$  & $-1$   & slip ($n^{*}{=}3$) \\
I  & $78$   & $0.300$ & $1$  & $-100$ & optimistic ($n^{*}{=}3$) \\
R1 & $78$   & $0.285$ & $1$  & $-165$ & ridge ($n^{*}{=}31$) \\
R2 & $80$   & $0.290$ & $1$  & $-260$ & ridge ($n^{*}{=}6$) \\
\addlinespace
\multicolumn{6}{@{}l}{\itshape crossing the limit at an ordinary interval ($L = 500$\,B):}\\
W  & $500$  & $2.0$   & $2$  & $0$    & at the limit, exact \\
X  & $500$  & $1.9$   & $2$  & $-400$ & unsound ($n^{*}{=}4$) \\
\bottomrule
\end{tabular}
\end{table}

\section{Discussion}
\label{sec:discussion}

The two use cases completed differently, and the difference is the
framework's class structure at work. On the TDM-PON use case the framework
delivered the dynamic contract class in full: a contract sound in all
sixteen examined cells and tight to about $1.2\,\mu$s, carrying its own
disposition instant per direction, and holding through imprecise
actuation, $200\times$ clock drift, and coexisting tenants---every claim
measured on the oracle (Section~\ref{sec:eval}). On the 5G--TSN bridge the
framework delivered the static contract class to its load-bearing step,
with the scheduling term exact against the kernel across the sweep and its
validity limit in closed form (Section~\ref{sec:tsn}). We keep the
evidential asymmetry explicit: the dynamic claims are measured, while the
static use case is kernel-referenced and its final verification remains
open.

Across both use cases, and from two different Cards, the framework's
reasoning role was reproduced. The handler, implemented with an LLM, acted
as a constructor under verification: it built the network models, surfaced
the Cards' defects, and ranked the discrete mechanisms
consistently across two LLM families (Section~\ref{subsec:repro},
Table~\ref{tab:brain}). The one task the architecture withholds from the
LLM is computing the guarantee, and the ablation measured why. Asked to
produce the worst-case delay directly, five of six draws were
optimistically unsound, and the error sat exactly where the physics is
tight, at the FBA-QF grant-quantization edge.
Our synthesis was designed to avoid that path. The kernel holds the
numeric authority and the oracle the authority to admit, and the handler
calls both as tools, so no probabilistic output enters the contract. 

The oracle's role differs by contract class, and the distinction organizes
what this evaluation establishes. On the dynamic use case the oracle is a
discoverer: the safety criterion itself had no a-priori algebraic form and
was established by observing reconfigurations. In that role
the handler drove the oracle to locate what no steady-state model
sees---the disposition instants. On the static use case the oracle is an auditor:
the value already sits in the kernel, and the oracle's job is to verify
it. That is the seat we left open on the 5G--TSN bridge, at the cost
Section~\ref{sec:tsn} states. Either way the contracts are verified, not
certified. Soundness is a
property of the checked output, not of the LLM that proposed it. The
oracle is finite, and catching omissions does not prove none remain. Its
fidelity bounds the framework's validity, and a structural reality absent from both the Card and the oracle is beyond any method. The oracle seat
accepts any fidelity, so swapping in a testbed or hardware-in-the-loop is
the deployment-validation path.

We claim nothing beyond the searched space. Over the Card's mechanisms,
enumeration with exact feasibility checks is complete, and the framework
is not a combinatorial optimizer. Where the constraint structure is
already algebraic, exact programs do produce configurations directly:
gate-schedule integer programs and SMT encodings among
them~\cite{tsn-ilp}. Such a program is a tool the framework can call, not
a rival. It occupies the kernel's seat at step~3, as a multi-queue kernel
that returns a schedule meeting the requirement rather than a value to
check, and the oracle still verifies what it returns. Two things bound its
use. Someone must build its input, the constraint structure and the
scenario, which is the handler's work here and is what we automate. And it
solves a static problem: it does not re-solve per announced regime, and
where it is fast enough to run online the lead the client commits would
have to cover its solve time. Either way it is one tool at one step, and
the framework is not its tools. That pattern sits earlier in the design pipeline than any solver: where a
constraint has no a-priori algebraic form, discover it once and hold it in
closed form thereafter. The transient-safety criterion followed that route on
the oracle side (Appendix~\ref{app:safety}), the exactness limit of the
single-drain form followed it on the kernel side (Section~\ref{sec:tsn}), and
the slotted kernel itself is the closed form that replaces a per-substrate
simulation of the grant beat (Appendix~\ref{app:comb}). Each time, a relation that had to be rediscovered
by simulation becomes one that can be stated once and checked thereafter,
and a stated criterion is what an automated handler carries from one
domain to the next.

Three steps follow from here. First, composition of the transient. The
standards compose per-domain bounds end to end, but when neighbouring
domains reconfigure together their safe apply sets must themselves
compose, and formalizing that joint safety along a path is, in our
judgement, the most consequential open step---the single-domain criterion
of Appendix~\ref{app:safety} is its foundation. Second, machine-certified
primitives. A kernel extended to evaluate return and the bounded
transient directly would retire the oracle from that check, making the
two-window criterion the first entry in a library of certified
blocks---scheduling, serialization, reconfiguration---from which the
handler selects rather than writes. Third, the dynamic upgrade of the
5G--TSN interface. It is static today, its changeovers re-registrations;
when announced transitions replace recomputation and reinstallation, the
machinery it needs is exactly the dynamic contract class, the harder case
this framework was built for. That step needs a packet-level 5G substrate
in the oracle seat.

\section{Conclusions}
\label{sec:conclusion}

We presented a framework that synthesizes the assume--guarantee contract a
deterministic cross-domain boundary needs. For the dynamic class, the
synthesis produced the contract itself: for each declared regime, the
configuration that meets the requirement and the disposition instant at
which to apply it. Standardized end-to-end composition consumes the
guarantee that results, and derives neither of its parts. The static,
registration-time class followed as the same workflow, ending at its
steady-state step. 
The architecture rests on a measured division of labor. A large language
model (LLM) implements the framework's two reasoning modules, the agent
and the handler: it constructed the network models, surfaced the Cards'
defects, and ran the workflow reproducibly across two LLM families and two
substrates. By design, the LLM holds no numeric authority. It drives two
passive tools to evaluate what it builds. An extended network-calculus kernel computes the worst-case metric for a
periodic arrival pattern served by a periodic schedule, exactly where the
two are commensurate and soundly elsewhere. An independent oracle, here a packet-level
simulator, locates what no steady-state model can express---the safe
disposition instant $t^{*}$, against the criterion we developed for
it---and verifies every contract before admission.
On uplink 5G fronthaul over a TDM-PON, the synthesized dynamic contract
was sound across the evaluated grid of arrival sizes, schemes, and line
rates, and tight to about $1.2\,\mu$s. It held the $100$-$\mu$s
requirement that reactive PON scheduling cannot, at up to 3.5 times the bandwidth efficiency a static contract over the same declared set would reach, and stayed robust under measured clock drift, late announcements, and coexisting tenancy. On the
5G--TSN bridge, the same workflow synthesized the static class the
standardized interface defines: the scheduling term exact against the
kernel across the industrial classes, its validity limit in closed form. A
single framework thus synthesized one interface of each class on two 
different carriers, and the reproducibility draws showed the construction
traveling from one Card to the other.
Composing safe apply sets across domain boundaries, so that neighbouring
domains may reconfigure together without either breaking its own contract,
remains the most consequential open problem.


\renewcommand{\appendixname}{Appendices}
\appendix

\subsection{The queueing term for two periodic processes}
\label{app:comb}

This appendix develops the queueing term the kernel returns---the value 
$H(\alpha,\theta)$ of Section~\ref{subsec:decomp}---under the service model our use cases share, where a scheduled, slotted service meets a periodic arrival pattern at the queue. Two assumptions hold throughout, and the final
paragraph revisits both. The traffic arrives in a periodic \emph{pattern}:
a fixed sequence of arrivals of $L$ bytes each, repeating with period
$T_{\text{pat}}$ and containing $n_{\text{a}}$ arrivals. The carrier
serves it on a periodic schedule $\theta$, one service instant every
$T_{\text{ser}}$, each carrying at most $G$ bytes. The arrival pattern need not be uniform. The service may follow a pattern too---a TSN gate schedule with several windows per cycle is one---and the
construction is similar. We keep the service uniform here, as both use cases have it. The fronthaul arrival of Section~\ref{sec:cti} is fourteen arrivals
in a $500\,\mu$s slot, thirteen spaced $35.677\,\mu$s and one
$36.198\,\mu$s, the longer symbol carrying the extended cyclic prefix so
that the slot closes on an integer count of the base clock. Where the
arrival pattern is uniform, $T_{\text{pat}} = T_{\text{arr}}$ and
$n_{\text{a}} = 1$.

\begin{figure}[t]
  \centering
  \includegraphics[width=\columnwidth]{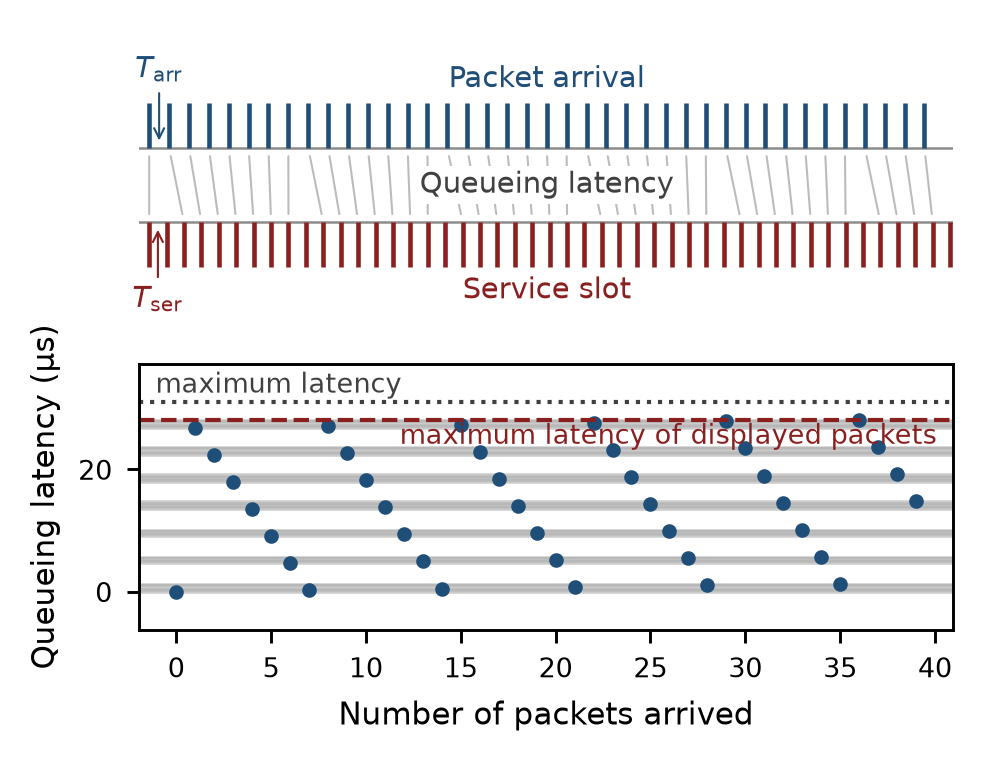}
  \caption{The beat of two combs. (a) The arrival pattern (5G symbols:
  $36.198\,\mu$s for the extended-cyclic-prefix symbol and $35.677\,\mu$s  for the other thirteen, closing the $500\,\mu$s slot exactly) against   the service comb ($T_{\text{ser}} = 31.25\,\mu$s, FBA-QF): each packet   waits for the next service instant. (b) The per-packet wait $w_k$; the peak-to-peak spread is the jitter.}
  \label{fig:comb}
\end{figure}

Each of these processes is a \emph{comb}: a train of instants, each
carrying a size. The arrival comb is the one on which packets enter and
the service comb the one on which they leave, and their interaction over
time is what we call the \emph{beat}. A packet that arrives waits for the
next instant of the service comb, and that wait is the queueing delay.
Because the two combs run at different periods, the wait changes from one
packet to the next as their relative phase advances
(Fig.~\ref{fig:comb}a). Standard rate--latency composition discards this
discrete structure~\cite{nc}. In this appendix we develop the model that
keeps it: the beat itself, its accumulation-aware grant-edge bound, and
the rate-aware window refinement---the two forms our use cases deploy.

Take first a uniform comb. The wait of the $k$-th packet is
$w_{k} = T_{\text{ser}}\lceil k\,T_{\text{arr}}/T_{\text{ser}}\rceil
- k\,T_{\text{arr}}$. We denote by $g = \gcd(T_{\text{arr}}, T_{\text{ser}})$
the base period the two combs share, and
$q = (T_{\text{arr}} \bmod T_{\text{ser}})/g$ for the number of comb
positions the phase advances per packet. Over a long run,
\begin{equation}
\begin{split}
Q = T_{\text{ser}}/g, \enspace
w_k \in \{0,\, g,\, \dots,\, (Q-1)\,g\}, \\
w_{k+1} = (w_k - q\,g) \bmod T_{\text{ser}} .
\end{split}
\label{eq:comb}
\end{equation}
The wait takes one of $Q$ distinct values, spaced by the base period $g$,
and it walks. From one packet to the next it steps down by $q\,g$. Where
that would take it below zero the modulo adds $T_{\text{ser}}$, so the wait
rises by $T_{\text{ser}} - q\,g$ instead.
The packet-to-packet variation is therefore at most
$\max(q\,g,\ T_{\text{ser}} - q\,g)$---on the fronthaul comb the wrap
dominates, at $26.8\,\mu$s against a $4.4\,\mu$s step down---and the peak-to-peak spread across the cycle, which is the \emph{jitter} this queue adds, is $(Q-1)\,g$. A non-uniform pattern
gives the same structure with a per-gap step. The wait walks down by the
phase advance of each inter-arrival gap and wraps the same way, and the sequence still repeats over the hyperperiod. Fig.~\ref{fig:comb} shows the beat for the fronthaul comb.

The handler carries this whole set---the beat of the two combs---rather
than collapsing it to a single worst-case scalar. Keeping the whole set is also what lets the kernel serve other QoS attributes. The maximum over the set is the worst-case delay, its spread is the jitter, and the backlog the same service curve bounds gives the buffer the element needs. Collapsing the set to a single scalar would discard all but the first. The varying wait also makes the served stream no longer
periodic. Successive packets leave the queue having waited different amounts, and the spread of those waits is the jitter (Fig.~\ref{fig:comb}b). That departing stream is what a kernel call returns alongside the delay term (Eq.~(\ref{eq:kernelcall})). In a chain, the beat at one element defines
the arrival comb at the next, and the jitter this queue adds is the
irregularity the next element must absorb.

Beyond this deterministic beat, real traffic deviates: the client's
emissions stray from their nominal instants by the jitter it declares,
the bounded clock error shifts the phase, and service is not perfectly
punctual---the uncontrolled parameters $\Phi$ of
Section~\ref{subsec:overview}. Treating these as strictly deterministic
understates the tail. Our kernel therefore does not stop at the
deterministic beat: it carries a bounded envelope on the total
deviation~\cite{snc} and propagates it into the queueing term, so what
$H$ returns is a deterministic bound with a stated stochastic tail rather
than a deterministic bound alone. The width of that envelope is a
physical trade-off: a tighter one is more efficient but tolerates less
deviation, while a looser one wastes capacity but absorbs more. The
framework makes the choice explicit, so that it is deliberate rather than
buried in a generic safety margin.

The pure wait $w_k$ covers only the case in which each arrival is served
whole at the next service instant. In general $L$ and $G$ do not stand in
that relation, and arrivals accumulate. We extend the beat to this
accumulation to obtain the kernel \emph{edge form}. By arrival $k$ the queue has received $(k+1)L$ bytes, which
take $\lceil (k+1)L/G \rceil$ service instants to clear, so arrival $k$
departs at that instant---the grant edge---and waits 
\begin{equation}
\begin{split}
h_{\text{edge}} = \max_{0 \le k < n}\,
\left[\, \lceil (k+1)L/G \rceil\, T_{\text{ser}} - t_{k}
\,\right],
\\
n = n_{\text{a}}\,\mathrm{lcm}(T_{\text{pat}},T_{\text{ser}})
/ T_{\text{pat}},
\label{eq:hedge}
\end{split}
\end{equation}
where $t_{k}$ is the instant of the $k$-th arrival within the
hyperperiod, $t_{0} = 0$, and the maximum runs over the $n$
arrivals that hyperperiod holds. The hyperperiod is the least period common to the arrival pattern and the service comb. For a uniform comb
$t_{k} = k\,T_{\text{arr}}$ and $n$ reduces to
$\mathrm{lcm}(T_{\text{arr}},T_{\text{ser}})/T_{\text{arr}}$. Where $T_{ser}$ divides $T_{pat}$ as it does for the fourteen 5G symbols of one 
$500\,\mu$s slot, the hyperperiod is the pattern period and $n=n_a$.
Because every period is an integer multiple of the base clock, the hyperperiod is finite and the maximum over it is exact. Nancy~\cite{nancy} evaluates the same model from the service curves and matches
Eq.~(\ref{eq:hedge}) at base-clock resolution (Section~\ref{sec:eval}),
which validates the construction rather than the physics.

The $h_{\text{edge}}$ form is rate-blind: it charges the unit its full
wait to the edge, as if the unit left the queue whole at that instant. The form has two reductions. The pure wait $w_k$ is the no-accumulation case developed above, where each arrival is served whole at the next instant. Eq.~(\ref{eq:hedge}) extends it to accumulation, and takes the conservative convention for an arrival that falls exactly on a service instant. The single-drain form of Section~\ref{sec:tsn}, one burst per interval, is the other. The closed form of Eq.~(\ref{eq:hedge}) has a
validity limit, reached when the service period approaches the arrival
period, and our second synthesis, the 5G--TSN bridge use case, develops it (Section~\ref{sec:tsn}).

The grant-edge form treats a service instant as a point. In reality it is
a window: a grant of $G$ bytes occupies the line for $G/R$ at line rate
$R$, and the unit's bytes leave throughout that window rather than at its
edge. We therefore extend the model a second time, to a rate-aware,
whole-packet service curve, and the kernel's bound of that curve is the
kernel \emph{window form} $h_{\text{win}}$, named for the time window the grant occupies. Unlike $h_{\text{edge}}$, it is obtained by
evaluating the curve rather than in closed form. The difference
$h_{\text{edge}} - h_{\text{win}}$ is the \emph{drain credit}: the part
of the edge-charged wait the unit spends already draining onto the line.
The credit scales with the grant duration, halving when the line rate
doubles, so the line rate enters the bound only here
(Section~\ref{sec:eval} measures it).

Soundness turns on one packetization rule: serialization is never folded
into the service \emph{rate}, the naive rate--latency curve a recent
critique identifies as unsound for this kind of
service~\cite{nc-packetization}. It is accounted for exactly once, and
where depends on the form. Under $h_{\text{edge}}$ the kernel returns the
grant wait alone, and serialization is the separate $d^{\text{trans}}$
term of Eq.~(\ref{eq:decomp}). Under $h_{\text{win}}$ the bound already
credits the unit draining across the grant, so the handler must not add
$d^{\text{trans}}$ again. The 5G fronthaul over PON use case uses the latter
(Section~\ref{subsec:cti-comb}), where $h_{\text{win}}$ is the bound of
record. This is bookkeeping internal to the construction, not a change to
the physical delay.

Two things here are our contributions: the accumulation-aware closed form
$h_{\text{edge}}$, and the rate-aware refinement $h_{\text{win}}$ with
the drain credit that separates them. Standard network calculus supplies
the service-curve machinery beneath them~\cite{nc}, and Nancy~\cite{nancy}
evaluates the curves. What we add is the construction that keeps the two
combs' interaction instead of averaging it away. We implemented both
forms as extensions of the framework's kernel---comb-construction and
evaluation layers above the unmodified Nancy engine. So for an arrival
pattern $\alpha$ ($L$ bytes per arrival, $n_{\text{a}}$ arrivals every
$T_{\text{pat}}$) served by a periodic schedule $\theta$ ($G$ bytes every
$T_{\text{ser}}$), the kernel returns both $h_{\text{edge}}$ and
$h_{\text{win}}$. This gives the handler a more accurate tool for
modeling that queue than a rate--latency envelope gives it. Note that
the handler computes neither form itself: it maps the carrier's Card's
physical parameters onto the model and calls the extended kernel.

The proposed synthesis framework is not confined to this exact base-clock
evaluation. A non-periodic arrival, a different discipline, or a path of
several queues relaxes the two assumptions above and triggers the
multi-element case of Section~\ref{subsec:decomp}, where network calculus
composes the service curves~\cite{nc}. What the framework adds rides
strictly on top of that standard composition---retaining the phase-aware
beat, the bounded-stochastic tail, and the reconfiguration transient of
Appendix~\ref{app:safety}---dynamics that a standard rate--latency
abstraction discards. The same beat governs both use cases: the TDM-PON
grants against the fronthaul comb (Section~\ref{sec:cti}) and the
configured 5G grant against the TSN gate schedule
(Section~\ref{sec:tsn}).

\subsection{Reconfiguration safety}
\label{app:safety}

This appendix defines the \emph{safety} criterion that step~4 evaluates and the disposition instant it yields. A reconfiguration is the transition between two \emph{operating points}, an operating point being a regime together with the configuration that serves it. We also call it a \emph{pair}. A reconfiguration spans both planes: the control plane carries the change, and the data plane carries its consequence in the queue. Each could in principle be modeled analytically---the actuation where its delay model is known, the transient by a kernel extended to evaluate it. Here neither is, so the oracle stands in for both, and the symbols below are those of the live network. Note that the criterion we develop here stands on its own and applies outside this synthesis. Following the paper's convention, a prime marks the previous
and a bare symbol the new. The previous operating point
$(\alpha',\theta')$ is in force. The client announces that the new regime
$\alpha$ takes effect at the onset $\tau$, the announcement arriving at
latest $\tau-\lambda_{c}$, and commits to a minimum spacing $T_c$ between
successive onsets (Section~\ref{subsec:overview}). The carrier answers
with the configuration $\theta$ its mapping $\alpha \mapsto \theta$
returns (step~3, verified at the start of step~4) and applies it at an
instant $t$ that it schedules. All instants are read on the shared clock (with its inaccuracy covered in $\Phi$).

We write $O(\alpha,\theta,\rho_0)$ for a pair's \emph{settled bound}: the
steady-state worst-case metric of that operating point over the
contracted span. The registration parameter is held at $\rho_0$ throughout
and suppressed from here on. Eq.~(\ref{eq:verify}) aligns the steady-state performance of the oracle with the analytic $\Delta$ the kernel's
terms assemble. 

Admissibility, the kernel's verdict at step~3 and its verification at the
start of step~4, is stated on a pair
\begin{equation}
\theta \text{ admits } \alpha \iff O(\alpha,\theta)\le\mathcal{D}_\alpha\label{eq:admit}
\end{equation}
a settled bound being finite only where the pair is \emph{stable}, so
Eq.~(\ref{eq:admit}) carries stability implicitly, as
Eqs.~(\ref{eq:feasible}) and~(\ref{eq:verify}) do. This is
Eq.~(\ref{eq:feasible}) under a shorter name, and every configuration the mapping returns admits its regime by construction.

The same symbol orders regimes and configurations, both through
admission:
\begin{equation}
\begin{split}
\alpha' \preceq \alpha &\iff \forall\,\vartheta:\ \vartheta \text{ admits } \alpha \implies \vartheta \text{ admits } \alpha' \\
\theta' \preceq \theta &\iff \forall\,\alpha:\ \theta' \text{ admits } \alpha \implies \theta \text{ admits } \alpha
\end{split}
\label{eq:order}
\end{equation}

We say $\alpha$ is \emph{heavier} than $\alpha'$, and $\alpha'$
\emph{lighter} than $\alpha$, when $\alpha'\preceq\alpha$: every
configuration serving the heavier regime serves the lighter one too, which
for arrival envelopes is pointwise dominance. Likewise $\theta'$ is
\emph{weaker} than $\theta$ when everything the weaker configuration
serves the stronger serves as well. Neither ordering is total. Both orderings are defined through admission, which carries each regime's own requirement. A regime with a tighter requirement can therefore be heavier at the same load, and two regimes can be incomparable because each is harder in a different respect. Heavier and lighter therefore do not reduce to a single scale in general. On
both our use cases the declared set does reduce to one, which is what lets
the handler test only the extreme transitions in each direction
(Section~\ref{subsec:oracle}). Both endpoints of a
transition are admissible, which the criterion presupposes. What it adds
is a demand on the interval between them, and on the state in which it leaves the second.

\begin{figure}[!htbp]
  \centering
  \includegraphics[width=\columnwidth]{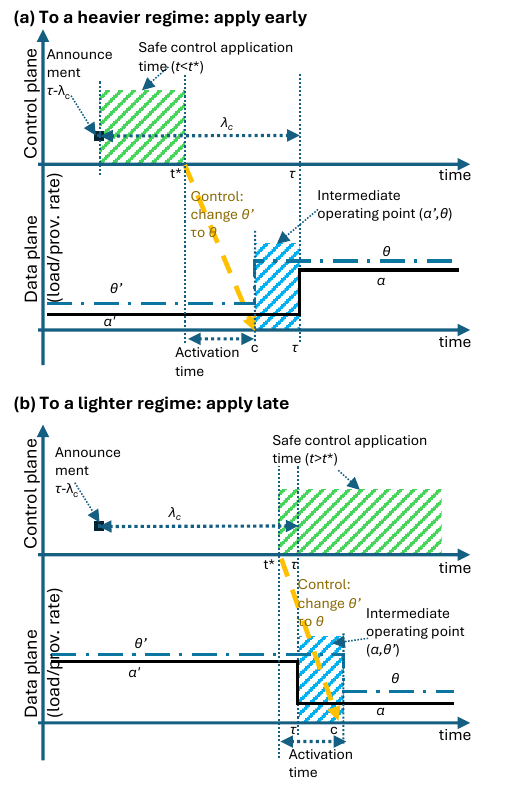}
\caption{A reconfiguration for a transition to a
heavier regime (a) and to a lighter one (b), shown as a load step. In the
data plane the arrival steps at the onset $\tau$ and the provisioned
service rate at the effective instant $c$; the carrier applies the change
at $t$ on the control plane, and actuation carries it to $c$. The intermediate interval between $c$ and $\tau$, where one regime
is served under the other's configuration, is shaded. The safe apply set lies on one
side of the disposition instant $t^{*}$---early to a heavier regime, late
to a lighter one---and $\lambda_c$ is the room the announcement leaves to
act.}
  \label{fig:disposition}
\end{figure}

The onset $\tau$ is the instant the regime changes. The change itself
crosses two planes: the carrier applies $\theta$ at the control-plane
instant $t$, and the applied configuration becomes \emph{effective} in
the data plane at an instant $c$. Two things separate the two. The
carrier's actuation may have its own control cycle---a PON
applies a grant change at a downstream frame boundary, so $c$ lands on a
$125\,\mu$s grid. Also within that control cycle the actuation delay
varies, by the control-plane pipeline, and the shared
clock's bounded error, the uncontrolled parameters $\Phi$ of
Section~\ref{subsec:overview}. Neither is under the synthesis's control,
so $c$ cannot be placed exactly at $\tau$, and the synthesis does not try to. It lands the effective change on one deliberate side of the onset,
\emph{early} ($c<\tau$) or \emph{late} ($c>\tau$). The \emph{disposition
instant} $t^{*}$ is the control-plane edge that guarantees the chosen
side: applying at $t\le t^{*}$ keeps $c$ before $\tau$, and applying at
$t\ge t^{*}$ keeps it after. Which side is chosen is set by the direction
of the regime change, and $t^{*}$ is pinned by reconfiguration tests, so that
the control cycle and its variation enter it by measurement rather than by
assumption. Both are developed below, and Fig.~\ref{fig:disposition}
shows the arrangement on both sides.

Either way the data plane traverses an \emph{intermediate pair}, the
queue holding one regime's traffic while served under the other's
configuration. The run decomposes into three settled pairs in sequence:
\begin{equation}
\begin{split}
\text{early}\ (c<\tau):\;
(\alpha',\theta')\big|_{[\tau-T_c,\,c)},\;
(\alpha',\theta)\big|_{[c,\,\tau)},\;
(\alpha,\theta)\big|_{[\tau,\,\tau+T_c]}, \\[2pt]
\text{late}\ (c>\tau):\;
(\alpha',\theta')\big|_{[\tau-T_c,\,\tau)},\;
(\alpha,\theta')\big|_{[\tau,\,c)},\;
(\alpha,\theta)\big|_{[c,\,\tau+T_c]}.
\end{split}
\label{eq:pairs}
\end{equation}
The transition run of Section~\ref{subsec:oracle},
$O(\alpha' \xrightarrow{\tau} \alpha,\; \theta' \xrightarrow{t} \theta)$,
is measured over this sequence. The carrier acts at $t$ on the
control plane, actuation carries the change to $c$ in the data plane, and
the middle interval of Eq.~(\ref{eq:pairs}) is the gap between $c$ and
the onset $\tau$---an interval the carrier shapes only through its choice
of $t$. The intermediate pair $(\alpha',\theta)\big|_{[c,\,\tau)}$ or $(\alpha,\theta')\big|_{[\tau,\,c)}$ may itself be admissible, and on the chosen side it is, by an inheritance shown below. 
But admissibility speaks about a settled run, and the intermediate interval is not settled: the latency can rise above both settled bounds, and that excursion is what the criterion must bound. Two things drive it. One is backlog---what the previous pair left in the queue has to clear under the new one, and admissibility does not say it clears within the window the contract needs. The other is phase: changing $\theta$ moves the service comb relative to the arrival comb, so the beat of Appendix~\ref{app:comb} re-aligns and the wait pattern that follows differs from the one that preceded it. A settled analysis of either endpoint sees neither.
Network calculus, as the
kernel implements it, bounds each settled pair but does not
bound the transition between them~\cite{nc}. The synthesis must therefore
choose the side, the disposition instant $t^{*}$, and the configuration,
and the criterion defines when that choice is \emph{safe}. We state it
over the contracted span, whatever the span holds: $O$ is end to end, so
nothing below restricts it to a single queue.

Admissibility of the two settled pairs, Eq.~(\ref{eq:admit}), says
nothing about the transition, and stability is too weak a demand for the
transient itself. A transition that parks the backlog at a higher but bounded level is stable, and can even sit within the requirement. What it loses is the baseline. The next transition then starts from a queue the previous one left behind, and a change that would have been safe from the settled state is not safe from this one. One transition passes and the second fails for a reason the first created. The criterion is therefore stated on windows of the run, and it asks the run to settle at the new pair's own steady state before the next change may arrive. The client's commitment $T_c$ sizes the windows: successive onsets are at least $T_c$ apart, so the run around one onset spans the $T_c$ before it and the $T_c$ after. The pre-event window is $w_1=[\tau-T_c,\tau)$ and the post-event
window is $w_2=[\tau,\tau+T_c]$. Writing $O_{w}$ for the transition run's
worst-case metric restricted to a window $w$, and $O_{w}^{\text{end}}$
for the same metric over the final portion of $w$, an apply instant $t$ is \emph{safe} when
\begin{equation}
O_{w_1}\le \mathcal{D}_{\alpha'}, \quad
O_{w_2}\le \mathcal{D}_{\alpha}, \quad
O_{w_2}^{\text{end}}\le O(\alpha,\theta),
\label{eq:safety}
\end{equation}
all three taken on the run $O(\alpha' \xrightarrow{\tau} \alpha,\; \theta' \xrightarrow{t} \theta)$, started from the settled state of $(\alpha',\theta')$. A packet is judged against the requirement of the regime that emitted it, so the windows are indexed by arrival: traffic arriving before $\tau$ is $\alpha'$ and owes $\mathcal{D}_{\alpha'}$, traffic arriving at or after $\tau$ is $\alpha$ and owes $\mathcal{D}_\alpha$. We write $O\big(\alpha' \xrightarrow{\;\tau\;} \alpha,\; \theta' \xrightarrow{\;t\;} \theta\big) \models \mathrm{safe}(T_c)$ when Eq.~(\ref{eq:safety}) holds of that run.

The first two conditions are the \emph{bounded transient}. A change is expected to disturb the queue, and what the contract owes is not that nothing happens but that nothing exceeds the requirement in force while it does. The intermediate pair falls in whichever window carries its own traffic: applied early, $(\alpha',\theta)$ occupies $[c,\tau)$ inside $w_1$; applied late, $(\alpha,\theta')$ occupies $[\tau,c)$ inside $w_2$. Neither window ever needs the other's requirement. Judging $w_1$ against $\mathcal{D}_{\alpha'}$ rather than against the settled bound of $(\alpha',\theta')$ also keeps successive changes consistent: where onsets fall exactly $T_c$ apart, $w_1$ is the previous change's $w_2$, over which the previous change guarantees the requirement and nothing tighter.

The third condition is \emph{return}. By the end of $w_2$ the latency must be back under the new pair's own settled bound and stay there. The check is one-sided, since latency below the reference is an improvement. This is what puts $T_c$ in the criterion: return must complete before the next change may arrive. The final sub-window over which return is judged is a property of the substrate's settling granularity and is declared with the use case; on the fronthaul use case it is the last two PON frames.

The \emph{direction} of the change is the order of the two regimes:
$\alpha'\preceq\alpha$ is an increase, a transition to a heavier regime, $\alpha\preceq\alpha'$ a
decrease. The direction selects the side, because it decides which
intermediate pair of Eq.~(\ref{eq:pairs}) is admissible. For an increase,
landing early yields $(\alpha',\theta)$, and admissibility is inherited
directly from Eq.~(\ref{eq:order}): $\theta$ admits $\alpha$, and
$\alpha'\preceq\alpha$, so $\theta$ admits $\alpha'$. The intermediate
interval is over-provisioned and harmless. Landing late would yield
$(\alpha,\theta')$, the heavier regime on the weaker configuration, for which no such inheritance holds. That pair need not breach the requirement---the
backlog it builds may still drain in time---but nothing certifies it in
advance, and establishing it would mean evaluating every transient it
could start from. Choosing the inherited side costs a little efficiency
and buys a guarantee that holds without that search. An increase
therefore lands early. For a decrease the roles reverse by the same
argument, and the safe side is late. Both sides of that argument hold where the two regimes compare. Where they do not, neither side inherits, and the handler tests both.

The disposition instant, by contrast, cannot be derived. Landing on the
right side of the onset is necessary but not sufficient, on either side.
Early, $t^{*}$ must place $c$ before $\tau$ across the control cycle and
the actuation variation, and far enough before it that the intermediate
interval is long enough to matter. Late, $c>\tau$ leaves the backlog
admitted under $\alpha'$ still draining after the onset, and the effective
change must clear it as well. How far from the onset the change must land, in both directions, is what the step-4 tests pin by measurement. What makes a single instant enough is monotonicity on the
chosen side: applying an increase still earlier only lengthens an
over-provisioned interval, and applying a decrease still later only
lengthens a well-provisioned one, so the safe set is one-sided and
$t^{*}$ is its edge---the last safe instant early, the first safe instant
late. The carrier then places $t$ inside that set, with a margin
exceeding the $\Phi$-scale imprecision, measured on the PON use case in
Section~\ref{sec:eval}. For an increase it must act between the
announcement and the edge, $\tau-\lambda_c \;\le\; t \;\le\; t^{*}$,
which is non-empty exactly when $\tau-t^{*}\le\lambda_{c}$, the
feasibility check of Section~\ref{subsec:overview}. A decrease sets
$t\ge t^{*}$ and simply waits, consuming no lead. Because the transient depends on the previous operating point as well as the
new one, the handler evaluates the extreme transition the declared set admits
in each direction and adopts its worst case, so an individual transition needs
no analysis of its own. Finally, the direction
is a property of the transition, not of the new regime alone, so
evaluating the contract for $\alpha$ requires the previous operating
point as well, and the carrier adapter holds that memory.

For each announced change the synthesis therefore decides \emph{when},
the side and its instant $t^{*}$, and \emph{how much}, the configuration
$\theta$ actually deployed. The two are not independent. If no instant
satisfies Eq.~(\ref{eq:safety}) at the mapped configuration $\theta$,
that configuration is too lean to clear the inherited backlog inside
$w_2$, and the handler raises it to some $\theta^{+}$ with
$\theta\preceq\theta^{+}$ before testing again: provisioning above the
mapped value drains the transient faster and eases return. On the PON
use case no raise was needed at the declared spacing. The mapped
$\theta$, applied at the correct disposition instant, satisfied the
criterion in every scheme, rate, and direction, and provisioning above it
changed the transient's magnitude, never which instant is safe
(Section~\ref{sec:eval}). When the mapped configuration is safe at its
disposition instant, safety reduces to timing. Configurations above it
remain admitted as margin, robustness bought at bandwidth efficiency, a
trade-off Section~\ref{sec:eval} prices. 

We are explicit about the status of each element. Eq.~(\ref{eq:safety}) is a specification. It is stated without reference to the forwarding mechanism, and it does not care whether the span is slotted or how many queueing elements it holds. What is per-substrate is the evaluation. The statement can be used across any substrate. 
Because the transient has no a-priori algebraic form, the conditions were measured on the fronthaul use case, whose oracle supplies the control-plane behavior the criterion needs, and worst-cased over the declared set and both directions of load change. The 5G--TSN use case is a
static contract with no announced change, so it neither exercises the
criterion nor tests it. Return is judged at the end of $w_2$: the run must have come back and stayed there before the
next change may arrive. The settled mean and variance are diagnostics, not
the criterion. A
kernel extended to evaluate return and the bounded transient directly
would retire the oracle from this check, the first candidate for the
machine-certified primitives of Section~\ref{sec:discussion}.

One generic property closes the appendix:
\begin{equation}
\begin{split}
\mathrm{SAFE}(T_c) &= \{(\alpha,\theta)\in \mathrm{STABLE} : \forall\,\alpha'\in A,\ \exists\,t:\ \\
&\quad O\big(\alpha' \xrightarrow{\;\tau\;} \alpha,\;
\theta' \xrightarrow{\;t\;} \theta\big) \models
\mathrm{safe}(T_c)\},
\label{eq:sets}
\end{split}
\end{equation}
the safe set collecting the operating points reachable safely from every regime the declared set admits, each at the configuration the mapping assigns it. That configuration is admissible by construction, so only the destination pair carries the $\mathrm{STABLE}$ restriction. Then
$\mathrm{SAFE}(T_c)\subseteq\mathrm{STABLE}$, and
the inclusion can be strict: a point that is admissible but needs a raise
to $\theta^{+}$ is stable and not safe, since the raised configuration is
a different operating point. The two coincide exactly when no raise is
ever required. The safe set is also monotone in the spacing, since a
longer $T_c$ admits a longer settling allowance and a longer window in
which to return:
\begin{equation}
T_c \le T_c' \;\Longrightarrow\;
\mathrm{SAFE}(T_c) \subseteq \mathrm{SAFE}(T_c') .
\label{eq:monotone}
\end{equation}
A substrate that fails the criterion may therefore satisfy it under a
slower client. This is the reconfiguration analogue of the gap between
stability and schedulability, and whether a substrate realizes it is an
empirical question, asked per substrate and per spacing. For the PON
use case the answer at the declared $T_c$, measured in
Section~\ref{sec:eval}, is $\mathrm{SAFE}(T_c)=\mathrm{STABLE}$: at every
provisioning level down to the floor the mapped configuration sufficed,
and what set the disposition instant was a frame boundary rather than a provisioning level.


\bibliographystyle{IEEEtran}
\bibliography{Sections/main}


\end{document}